\documentclass[11pt,fleqn,letter]{article} %
\usepackage{graphicx}
\usepackage{latexsym} 
\usepackage{namedplus} 
\usepackage{hyperref}
\usepackage[svgnames]{xcolor}
\usepackage{amsmath}
\usepackage{amssymb}
\usepackage{bbold} 
\usepackage{color}
\hypersetup{colorlinks=true, 
citecolor=Blue, 
linkcolor=Blue, 
urlcolor=Blue}

\numberwithin{figure}{section} 
\numberwithin{table}{section} 
\numberwithin{equation}{section} 

\newcommand{\erf}{\mathop{\mathrm{erf}}} 
\newcommand{\vsp}{\vspace{0.2cm}}
\newcommand{\be}{\begin{equation}}
\newcommand{\ee}{\end{equation}}

\newtheorem{remark}{Remark}[section]
\newtheorem{defn}{Definition}[section] 

\begin{document}
\title{The mean field approach for populations of spiking neurons \footnote{This article is Chapter~6 of the Springer book ``Computational Modelling of the Brain. Modelling Approaches to Cells, Circuits and Networks". {\em Cite as:} La Camera, G. (2022). The Mean Field Approach for Populations of Spiking Neurons. In: Giugliano, M., Negrello, M., Linaro, D. (eds) Computational Modelling of the Brain. Advances in Experimental Medicine and Biology, vol 1359. Springer, Cham. \url{https://doi.org/10.1007/978-3-030-89439-9_6}.}}
\date{}
\author{\Large Giancarlo La Camera}
\maketitle

\vspace{-0.5cm}
\begin{center}
\small
Department of Neurobiology and Behavior, Stony Brook University

\vsp Program in Neuroscience, Stony Brook University

\vsp Center for Neural Circuit Dynamics, Stony Brook University

\vsp Stony Brook, NY, USA 

\vsp {\tt giancarlo.lacamera@stonybrook.edu}
\end{center}

\vsp \vsp \vsp
\begin{center} {\sc Abstract} \end{center}

Mean field theory is a device to analyze the collective behavior of a dynamical system comprising many interacting particles. The theory allows to reduce the behavior of the system to the properties of a handful of parameters. In neural circuits, these parameters are typically the firing rates of distinct, homogeneous subgroups of neurons. Knowledge of the firing rates under conditions of interest can reveal essential information on both the dynamics of neural circuits and the way they can subserve brain function. The goal of this chapter is to provide an elementary introduction to the mean field approach for populations of spiking neurons. We introduce the general idea in networks of binary neurons, starting from the most basic results and then generalizing to more relevant situations. This allows to derive the mean field equations in a simplified setting. We then derive the mean field equations for populations of integrate-and-fire neurons. An effort is made to derive the main equations of the theory using only elementary methods from calculus and probability theory. The chapter ends with a discussion of the assumptions of the theory and some of the consequences of violating those assumptions. This discussion includes an introduction to balanced and metastable networks, and a brief catalogue of successful applications of the mean field approach to the study of neural circuits.

\vsp \vsp \small
{\tt Keywords:} leaky integrate-and-fire neuron, binary neuron, logistic neuron, neural population, neural circuits, firing rate, asynchronous state, bistability, multistability, metastable dynamics. \normalsize

\newpage
\setcounter{tocdepth}{3} 
\addtolength{\parskip}{0.cm} 
\tableofcontents 
\section{Introduction} 
\label{sec:into}

The purpose of this chapter is to give an elementary introduction to mean field theory for populations of spiking neurons. Mean field theory is a conceptually simple, but far reaching method developed in physics to explain a wide range of phenomena, most notably to understand the nature of phase transitions \cite{Binney:1992tg,Le-Bellac:2004jj,Parisi:1998hp}. At its heart, it consists of neglecting fluctuations in the interaction between the units defining the system, and it can lead to qualitatively correct insights with relatively little effort. For example, a mean field assumption on the energy potential of a non-ideal gas leads quickly to the van der Waals equation of state \cite{Binney:1992tg}. Mean field results also have a weak dependence on the microscopic details of the system, promising to extract general principles that apply to large classes of seemingly unrelated models. 

As neural circuits of the brain comprise a large number of interconnected neurons, they are ideally suited to a mean field analysis. Very often, the goal is to capture properties of neural circuits that occur during {\em typical} behavior. Typical behaviors pertain to large networks and should not depend on the specific number of neurons (as long as this number is large), or the details about the neuron model, or the precise values of the synaptic weights. For this reason we are often interested in the properties averaged across the distribution of possible weights, and in the limit of infinite network size. Some important properties, such as the existence of a sharp phase transition, are only obtainable in this limit. In neuroscience, phase transitions are related, for example, to the existence of memory phases \cite{a89}, or to transitions between qualitatively different dynamical regimes \cite{Sanchez-Vives:2017ng}.

In this chapter we present the main ideas of the theory in a network of simplified neurons with probabilistic spiking. Including a probabilistic element allows to interpret the neural activities as random variables and to articulate the approach in a general language. All the main steps of the approach, together with its neural applications, are already available in this simple system, and can be grasped unencumbered by the technical difficulties that arise in networks of spiking neurons. When presenting the theory for integrate-and-fire neurons, an effort is made to eschew those difficulties and rely only on standard calculus and probability theory. The assumptions of the theory, and some possible departures from its predictions, are also discussed. Three important examples, bistable, metastable, and balanced networks are also briefly considered. 

We hope that this chapter will remove a gap in the existing literature by presenting an elementary introduction to the application of mean field theory to networks of spiking neurons. 

\section{Networks of binary neurons}
Consider a network of $N$ binary neurons $x_i \in \{ 0,1\}$, mutually connected by synapses $J_{ij}$ and receiving external input $I_{i,ext}$ coming from distant neurons in e.g. different brain areas. The input current to unit $i$ is therefore
\begin{equation} \label{eq:I}
I_i = \sum_{j \neq i}^N J_{ij} x_j + I_{i,ext},
\end{equation}
where the sum goes over all $N$ neurons $j$ except neuron $i$ itself. We assume that time proceeds in discrete time steps. At each time step, unit $x_i$ will emit a spike ($x_i=1$) if the input current is larger than a threshold $\theta$. For convenience and greater generality, it's best to assume a probabilistic process of spike emission. We assume that neuron $i$ will emit a spike with probability
\be \label{eq:Px}
p(x_i=1|I_i) \propto e^{\beta (I_i - \theta)},
\ee
where $\beta$ is a parameter that controls the level of stochasticity of the spike emission process. Note that, due to the terms $I_{i,ext}-\theta$ in Eq.~\ref{eq:Px}, the effect of the thresholds can be included in the external currents. Therefore, in most of the following, we set $\theta=0$ and keep the external currents constant. The following arguments generalize easily to the case of stochastic external currents.  

\begin{figure}[t]
\centering
\includegraphics[scale=0.4]{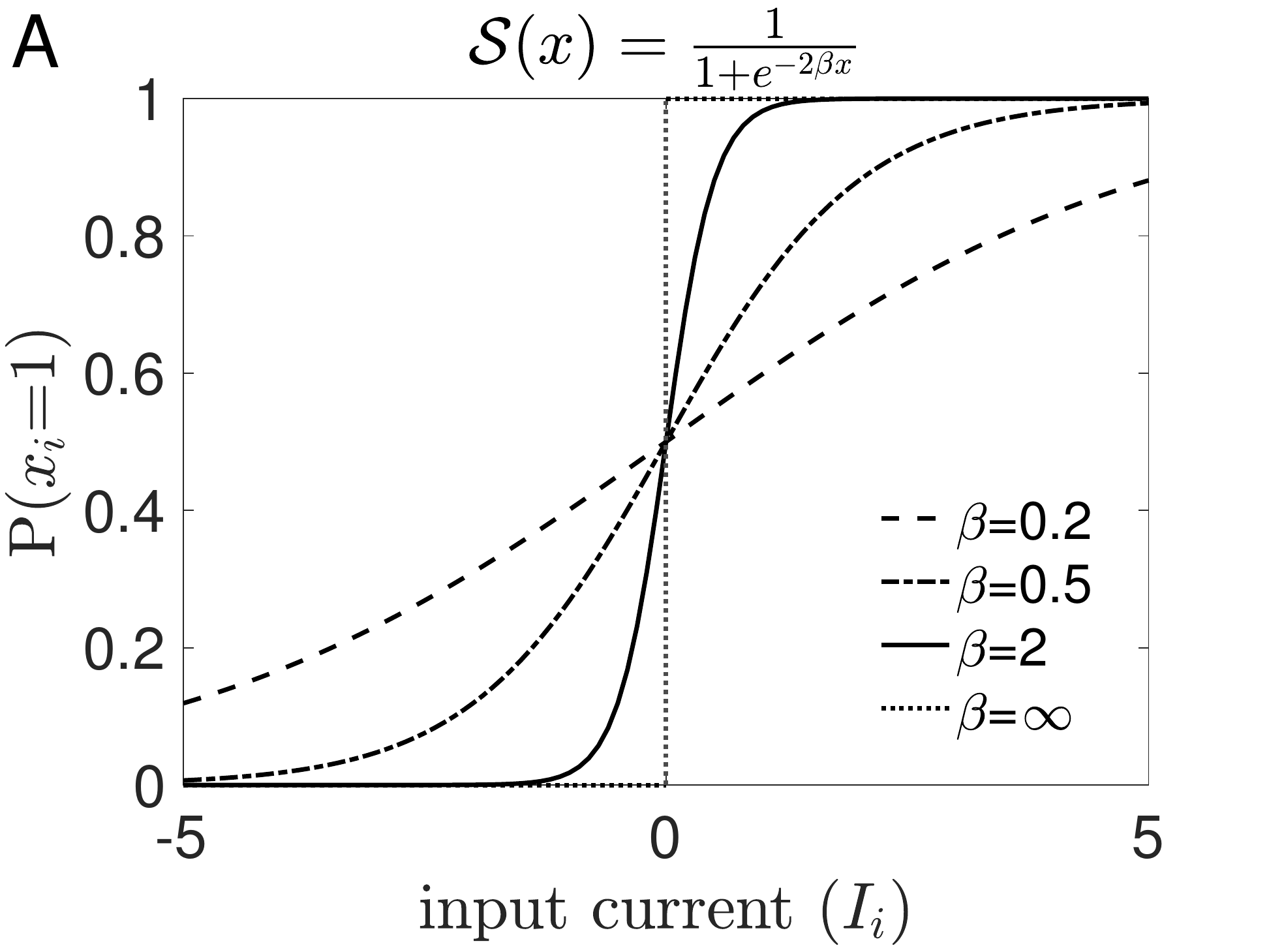}
\includegraphics[scale=0.4]{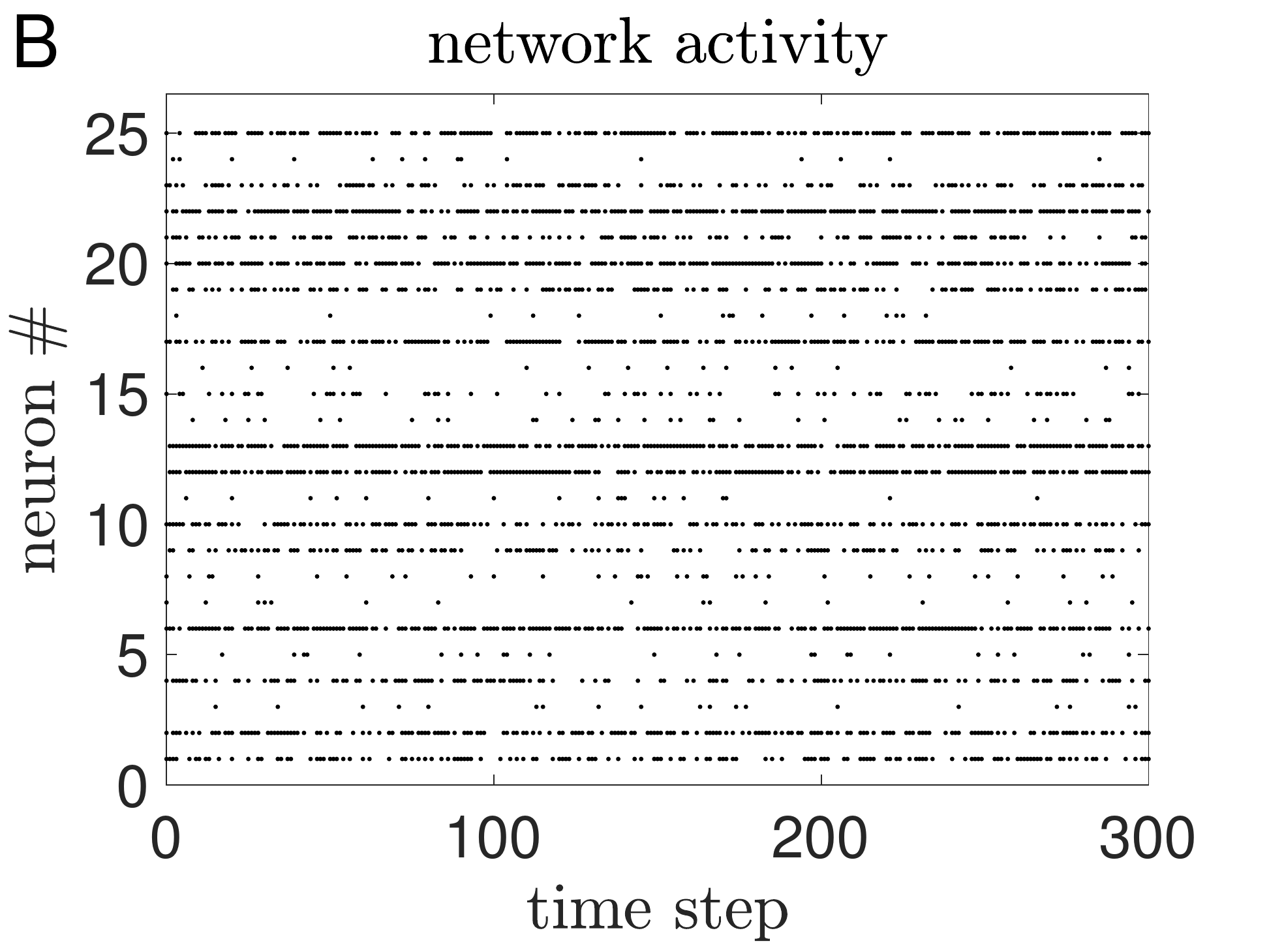}
\caption{\small {\bf A.} Plot of the logistic function $\mathcal S(x) = { 1 \over 1 + e^{-2 \beta x}}$ (Eq.~\ref{eq:Pspike}) as a function of $x=I_i$ for several values of $\beta$. For infinite $\beta$, the curve becomes the Heaviside function $\Theta(x)=0$ if $x<0$, $\Theta(x)=1$ if $x\geq 0$ (dotted line). {\bf B.} Spiking activity of a population of $1,000$ binary neurons (only $25$ shown; each line is a neuron, each dot is a spike). Here, $\beta=2$, the external currents and the synaptic weights were uniformly distributed across neurons: $I_{i,ext} = -1.1 u$, $J_{ij} = 2u/N$, where $u$ is a random variable uniformly distributed between zero and one: $u \sim \mathcal U(0,1)$. At each time step, all neurons are updated simultaneously, based on the value of their input current. }
\label{fig:logist}      
\end{figure}

Since the probability must be bounded by $1$, a convenient choice for $p(x)$ is the logistic function,
\be \label{eq:Pspike}
p(x_i=1|I_i) = { e^{\beta I_i} \over e^{\beta I_i} + e^{- \beta I_i}} = { 1 \over 1 + e^{- 2 \beta I_i}} \equiv \mathcal S(I_i),
\end{equation}
where $\mathcal S$ is the logistic function defined by the above equation. Normalization implies $p(x_i=0)=1-p(x_i=1)$, and therefore
\be
p(x_i=0|I_i) ={ e^{-\beta I_i} \over e^{\beta I_i} + e^{- \beta I_i}} = { e^{- 2 \beta I_i} \over 1 + e^{- 2 \beta I_i}}={ 1 \over 1 + e^{2 \beta I_i}}.
\ee
We can therefore write $p(x_i)$ compactly as
\be \label{eq:Pspike2}
p(x_i|I_i)=\mathcal S(I_i)^{x_i} \; (1-\mathcal S(I_i))^{1-x_i}, \quad x_i=\{0,1\}.
\ee
$\mathcal S(I_i)$ is plotted in Fig.~\ref{fig:logist}A for several values of $\beta$. Note that when $\beta \to \infty$, we retrieve the deterministic model of spike emission (see Fig.~\ref{fig:logist}A, dotted line). On the other hand, when $\beta \to 0$, $p(x_i=1)=0.5$, and the network becomes a population of independent neurons. Hence, $\beta$ also controls the degree of mutual influence among the neurons.\footnote{More precisely, when $\beta \to 0$ the neurons are still interacting, but random fluctuations in the spiking process completely overcome the impact of the other neurons.}   Because of Eq.~\ref{eq:Pspike}, we call this model the (binary) logistic neuron.

At every discrete time step, all neurons' activities are updated at the same time, according to the probabilistic rule Eq.~\ref{eq:Pspike2}.

%

\section{Characterization of neural activity}

The activity of a population of logistic neurons is shown Fig.~\ref{fig:logist}B for a given choice of parameters. This is an example of `raster plot', where each line is the spike train emitted by one neuron, and each dot is a spike time. We note that the activity of the network is stationary in the following sense: the {\em mean} input current $I_i$, and therefore the probability of emitting a spike given $I_i$, does not change with time. The activity still looks erratic, because the neurons probabilistically flip their states over time. This dynamical behavior is called the `asynchronous irregular' regime of cortical neurons \cite{av93,ab97,vs96,g00,Renart:2010pb}, to distinguish it from other collective behaviors such as global oscillations, regular spiking, bursting and so on (a full account of the possible dynamical behaviors of one relevant model can be found in refs. \citetext{bh99,b00}).

The asynchronous regime is therefore one in which the firing rates are constant in time, but the activities of the single neurons are uncorrelated and erratic, resembling a stochastic process. This regime is often observed in cortical circuits when recordings are made in behaving animals \cite{a95,hea96,compte03,London:2010ea}, and it can be reproduced also in networks of neurons via a number of mechanisms. We will say more about this later on. In the asynchronous irregular regime the average spike counts do not change, an observation that often motivates arguments of firing rate coding \cite{a95,London:2010ea}. Even so, firing rates do not completely characterize the dynamics of populations of neurons. The variability of the inter-spike intervals could also be of interest. Another relevant property is the temporal correlation of each neuron \cite{Joelving:2007ma,compte03} as well the pattern of pair-wise spike count correlations among different neurons \cite{Josic:2009uh,Doiron:2016qd}. In principle, all higher-order correlations among neurons would be of interest, although they are much harder to quantify \cite{Riehle:1997yg,Ohiorhenuan:2010rv,Gao:2017}.

Although mean field theory is `custom-made' to succeed in the asynchronous irregular regime, it can often provide information on the other aspects of the dynamics mentioned above. In this introductory account, we shall limit ourselves to the characterization of the asynchronous regime.

\subsection{Firing rate}

As pointed out in the previous section, when the general goal is to understand the aggregate, macroscopic behavior of populations of cortical neurons, collections of firing rates is a relevant place to start. Intuitively, the firing rate is a measure of the average activity of a neuron (or the whole network) based on the spike count, however there exists more than one definition of firing rate (see e.g. chapter~1 of ref. \cite{Dayan:2001fk}). For the logistic neuron, the firing rate of neuron $x_i$ (at any time $t$) can be defined with respect to the probability measure Eq.~\ref{eq:Pspike2} and is a function of $I_i(t)$:
\begin{eqnarray}
f_i(t) & = & \langle x_i(t) \rangle = 1 \times p(x_i(t)=1|I_i(t)) + 0 \times p(x_i(t)=0|I_i(t)) \\
& = & p(x_i(t)=1|I_i(t)) \\
& = & \mathcal S(I_i(t)). \label{eq:fix}
\end{eqnarray}
Here, the symbol $\langle \cdot \rangle$ is used for the average with respect to the distribution Eq.~\ref{eq:Pspike2}. Since $I_i(t) = \sum_{j \neq i} J_{ij} x_j(t)+I_{i,ext}$, $f_i$ is a function of all $x_j(t)$ with $j \neq i$. Since the latter are random variables, $f_i$ is also a random variable. We have averaged the spiking activity of unit $i$, but the result depends on the activities of the other neurons. Therefore, in some cases we further average $f_i(t)$ over the remaining $x_j$: 
\be \label{eq:fgen}
\langle f_i(t) \rangle_x = \langle \mathcal S(I_i(x(t))) \rangle_x = \sum_x \mathcal S(I_i(x(t))) \; p(x(t)),
\ee
where the vector $x \doteq \{x_1, ..., x_N\}$ does not contain $x_i$. We call this quantity the {\em average} firing rate to distinguish it from the firing rate $f_i(t)$. One could also be interested in higher moments of $f_i$ or, in general, in its probability distribution. We will see later that this probabilistic notion of firing rate is useful also in deterministic networks which are nevertheless capable of generating stochastic-like activity. Note that in a recurrent network, the $x_j$ will depend in turn on $\{ f_k\}$, in other words, Eq.~\ref{eq:fgen} is a {\em self-consistent} equation (more on this later). We now provide a few concrete examples that are relevant for the following. 

\subsubsection{Constant input}

If the neuron is probed by a constant input current $I_i$, then from Eq.~\ref{eq:fix} we simply have
\begin{equation} \label{eq:fI}
f_i = \mathcal S(I_i) = { 1 \over 1 + e^{-2 \beta I_i}}.
\end{equation}
This quantity is analogous to the frequency-current ($f$-$I$) curve in neurophysiology \cite{lgsf08}. In more general contexts this function goes under such names as `gain function', `transfer function' or `response function'. Fig.~\ref{fig:logist} shows that the response function of the logistic neuron is a sigmoidal function of the input current. Mean field theory reflects the generic properties of the response function such as its sigmoidal shape -- although its detailed shape may also qualitatively change the behavior of some networks \cite{md02,Kadmon2015-cb}. 

Note that different neurons may have different $f$-$I$ curves $\mathcal S_i$, in which case Eq.~\ref{eq:fI} is replaced by $f_i = \mathcal S_i(I_i)$, but the arguments given below proceed in much the same way. Also note that, since $f \in [0,1]$, $f$ values should be interpreted in units of maximal firing rate; for example, interpreting each time step as a time bin of $10$~ms, $f=0.1$ would correspond to a firing rate of $10$~spikes/s.

\subsubsection{Gaussian input current}

In mean field theory one considers the input current to be either a constant, as in Eq.~\ref{eq:fI}, or a Gaussian random variable $I_i(t)=I_i(z(t))$. At every step, $z(t)$ takes a random value according to a distribution $G(z)$ which we consider time-independent throughout this chapter. When the spiking activity of the neuron depends only on the current value of the input, as is the case of our logistic neuron, the firing rate is given by an average over the distribution of $z(t)$:
\be \label{eq:fgauss}
\langle f_i \rangle_z = \langle \mathcal S(I_i) \rangle_z = \int dz G(z) \; \mathcal S(I_i(z)).
\ee
Here we have suppressed the dependence on time due to our assumption of a stationary distribution $G(z)$. Note that, in a recurrent network, $z$ in turn depends on the activity of the network. We will see examples later on. 

\subsection{Measuring the firing rate} 

How do we measure, in practice, the firing rate of neurons? When the neural activity is stationary, i.e., the spiking probability does not change with time, the average firing rate can also be computed as the average spike count over time:
\be \label{eq:ftimeaver}
\langle f_i \rangle = \lim_{T \to \infty}  {1 \over T } \sum_{t=1}^T x_i(t) \approx {n_i(T) \over T},
\ee
where $n_i(T)$ is the number of spike emitted by neuron $i$ over a sufficiently long time $T$. Note that $\langle f_i \rangle$ gives the average firing rates of the neurons even as they will continuously flip their activity states (between spiking and non-spiking), as shown in Fig.~\ref{fig:logist}B. Since the spike trains are erratic, local temporal fluctuations of activity around the mean firing rates are expected, but they are suppressed by the dynamics of the network if the asynchronous state is stable. When all neurons in a population have the same mean firing rate, the latter can be estimated more accurately via an ensemble average, such as that defined in Appendix~\ref{app:EA}. The ensemble average also allows to measure the temporal modulations of firing rate in non-stationary situations.

\section{The mean field equations} \label{sec:mfmain}

The goal of mean field theory is to predict the behavior of our network and in particular how this behavior depends on its parameters. This is not an easy task, due to the interactions between the neurons. 

\vsp {\em The main idea of the mean field approximation is to replace the interaction between a neuron and its afferents with a mean field generated by the latter.}\vsp 

In other words, one assumes that the neurons in the network receive an input current equal to the mean input generated by their presynaptic neurons (in physics, where this approach was invented, atoms and elementary particles are under the effect of `fields', which explains the name `mean field'). The argument is as follows. One notes that the input current is a sum of $N$ random variables. If the individual variables are independent and $N$ is large, the central limit theorem tells us that the sum tends to follow a Gaussian distribution. Therefore we write: 
\begin{equation} \label{eq:I'}
I_i(t) = \sum_{j \neq i}^N J_{ij} x_j + I_{i,ext} \approx \langle I_i \rangle + \eta(t),
\end{equation}
where $\eta(t)$ is a temporally fluctuating Gaussian variable with stationary statistics (the extension to time-dependent processes will not be considered in this chapter). We later show how to include the effect of $\eta$ in our mean field approximation. But to start, we simply {\em assume that the fluctuations of $I_i$ can be neglected.} If the weights $J_{ij}$ are constants, this leads to the mean field approximation: 
\be \label{eq:meanI0}
I_i \approx \langle I_i \rangle = \sum_{j \neq i} J_{ij} \langle x_j \rangle + I_{i,ext} = \sum_{j \neq i} J_{ij} f_j + I_{i,ext},
\ee
where, to lighten the notation, by $f_j$ we mean the firing rate averaged over the activity of the whole network, $\langle f_j(x) \rangle_x$, see Eq.~\ref{eq:fgen}. Each neuron therefore experiences an input current which is equal to its mean. 
Replacing $I_i$ with its mean value into Eq.~\ref{eq:fI} we get
\begin{equation} \label{eq:meanf0}
\boxed{f_i = \mathcal S \left (\sum_{j \neq i} J_{ij} f_j + I_{i,ext} \right ), \quad i=1, ..., N.}
\end{equation}
This is our first example of mean field equations. They are a set of $N$ coupled equations for the firing rates $f_i$ of the $N$ neurons in our population. Comparison with Eq.~\ref{eq:fgen} shows that we have replaced $\langle \mathcal S(I) \rangle$ with $\mathcal S(\langle I \rangle)$:
\be \label{eq:meanfieldapprox}
\langle f_i \rangle = \langle \mathcal S( I_i ) \rangle = \mathcal S(\langle I_i \rangle).
\ee
Since $\mathcal S$ is a non-linear function, this relationship cannot be correct, in principle. The idea is that the input currents $I_i(x)$, as random variables, converge to their means $\langle I_i(x) \rangle$ in the limit $N \to \infty$, in which case $f_i \to \mathcal S(\langle I_i(x) \rangle)$. This procedure is basically an application of the law of large numbers to $I_i(x)$. This requires some care, an issue we consider in Sec.~\ref{sec:scaling}.

\vsp Note the following about Eq.~\ref{eq:meanf0}:

\begin{itemize}
\item spiking has disappeared and it has been replaced by smooth variables $f_i$;

\item the mean field equations are {\em self-consistent equations} in that the same mean firing rates appear on both the left and the right hand side of the equations.
\end{itemize}

The self-consistency requirement is due to the recurrent nature of the network, wherein the output of a neuron is also an input to all the other neurons. This is even more apparent if we write these equations in vectorial form, after defining the vector of firing rates $f=\{f_1, f_2, \dots, f_N\}$, the vector of external inputs $I_{ext}$, and the synaptic matrix ${\pmb J }$ (having elements $J_{ij}$ with $J_{ii}=0$): 
\be \label{eq:meanfvect0}
f ={\mathcal S} ({\pmb J } f +I_{ext}). 
\ee
In this equation, the vector $f$ is required to be the same on the left and right hand side, and for this reason it is called a `fixed point':
\begin{defn}[fixed points]
The self-consistent solutions of the mean field equations are called {\em fixed points} of the network's activity. 
\end{defn}
Depending on the nature of the model, there may be multiple fixed points, which may be stable or unstable. Often the aim is to build a model with fixed points having desired properties. We shall see examples later.

\vsp In summary, the mean field equations are self-consistent equations for the average firing rates, obtained under the hypothesis that we can replace the input to each neuron with its mean value.

\subsection{Solving the mean field equations}

One way to solve the mean field equations is to use a fictitious dynamics that converges to the solution. One convenient dynamics is the following:
\be \label{eq:meanIeq}
\left \{ \begin{array}{ll} 
f_j  =  \mathcal S(I_j) \\ 
\tau_I \dot I_i =  -I_i + \sum_{j\neq i}^N J_{ij} f_j + I_{i,ext}.
\end{array} \right.
\ee
At equilibrium, this system gives our mean field equations Eqs.~\ref{eq:meanI0}-\ref{eq:meanf0} for the pair $(I^*,f^*)$, where $f^* = \mathcal S(I^*(f^*))$ are the fixed points of this coupled system. 

The full dynamics of the pair $(I,f)$ would require closing an equation for the moments of $I$ as a function of $f$ (see e.g. \cite{bress09,Buice:2013ph} for examples of this kind of approach); however, the simplified dynamics \ref{eq:meanIeq} is effective at finding the fixed points of our network.


\begin{remark} -- {\em The model Eqs.~\ref{eq:meanIeq} is an example of `rate model' of Cowan-Wilson type \cite{wc72} and has been used in different contexts. For symmetric synaptic weights, it can be used as a `mean field version' of a stochastic network called the Boltzmann machine (which is closely related to our binary logistic network), see \cite{Hopfield:1984mu} and Ch.~7 of \cite{Dayan:2001fk}. For Gaussian random weights with zero mean and variance $g^2/N$, it has been analyzed to explore the ability of neural networks to produce chaotic dynamics in the firing rates \cite{Sompolinsky1988-tn}. In general, rates models are ad hoc descriptions of neural dynamics that can be derived as mean field approximations of microscopic models. Note that the same rate model can be interpreted as the mean field approximation of more than one microscopic description, see e.g. \cite{Cowan:2016ob,Chow:2020it} for recent reviews.}
\end{remark}

\subsection{Random weights} \label{sec:randomJ}

A highly relevant case is when the synaptic weights are random variables sampled from a given distribution. This is motivated by the fact that weights distributions in cortex are wide \cite{Buzsaki:2014pe}. As we are interested in the typical behavior of the network, we must average our quantities of interest over the distribution of synaptic weights. Importantly, {\em once the weights are sampled, they are kept fixed} (or `quenched'). It is said that they give rise to {\em quenched} noise, in contrast to {\em fast} noise emerging from the spiking dynamics of the neurons. Quenched noise is very important as it allows to include the effect of heterogeneities in the description of the collective behavior of neural circuits.

Also in this case, the mean field approximation assumes that we can replace the current with its mean:
\begin{equation} \label{eq:meanI1}
\langle \langle I_i \rangle \rangle = \sum_{j \neq i} \langle \langle J_{ij}  x_j \rangle \rangle + I_{i,ext},
\end{equation}
where we have used the symbol $\langle \langle \cdot \rangle \rangle$ to indicate an average with respect to the distribution of the weights and with respect to the distribution of the temporal values of the activities $x_j(t)$. We shall use the symbol $[ \cdot ]$ for the former and $\langle \cdot \rangle$ for the latter, so that
\be \label{eq:meanI2}
\langle \langle I_i \rangle \rangle = \sum_{j \neq i} [ J_{ij} ] \langle x_j \rangle + I_{i,ext} = J \sum_{j \neq i}  f_j + I_{i,ext},
\ee
where we have assumed the weights are independent samples from a distribution with mean $J$, and that the neural activities and the weights are uncorrelated variables. Note that the average $\langle x_j \rangle$ now depends on the distribution of the weights (this will be clearer in Sec.~\ref{sec:variance}), however we simply write $\langle x_j \rangle$ to simplify the notation.

Performing the mean field approximation, Eq.~\ref{eq:meanfieldapprox}, we get the mean field equations:
\begin{equation} \label{eq:meanf}
f_i = \mathcal S \left (J \sum_{j \neq i}^N f_j + I_{i,ext} \right ), \quad i=1,2,...,N.
\end{equation}

We defer an analysis of the approximations performed so far to a later section (Sec.~\ref{sec:validity}); first, we show how the theory can be used to make predictions on the network's activity. 

\begin{figure}[t]
\centering
\includegraphics[scale=.4]{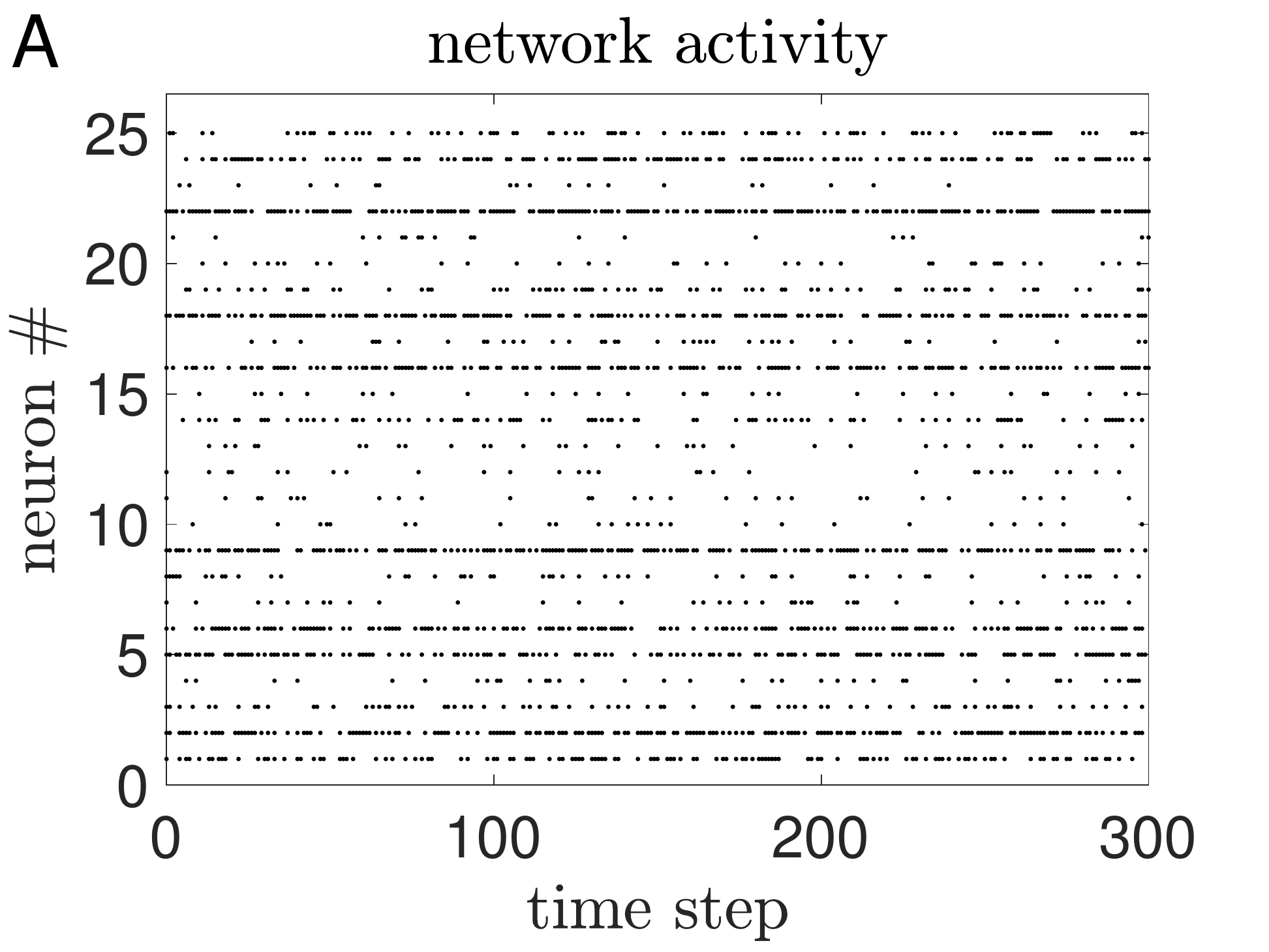} 
\includegraphics[scale=.4]{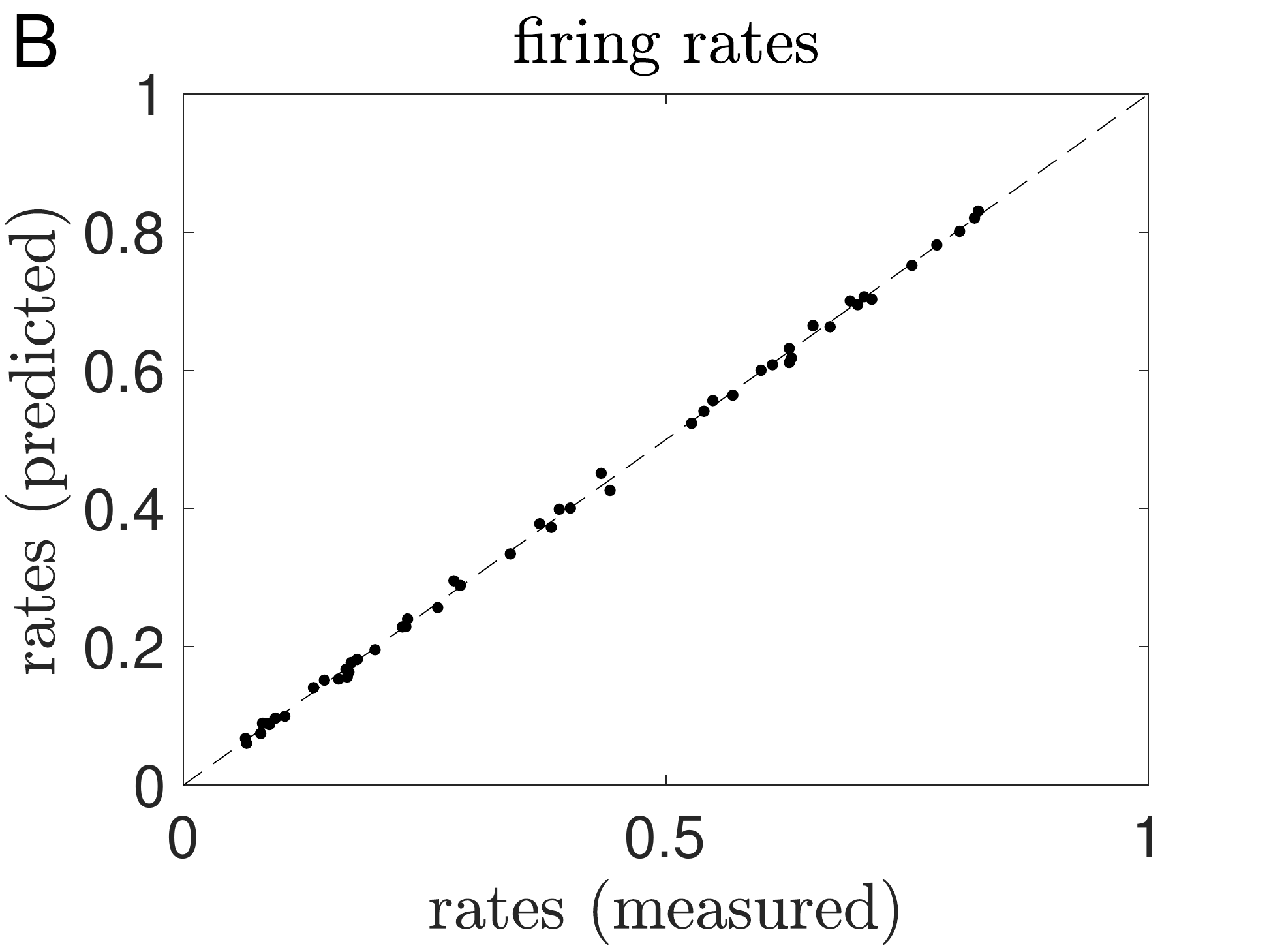}
\caption{Heterogeneous network with mean field predictions. {\bf A.} A raster plot from the network of Fig.~\ref{fig:logist}B (only 25 neurons shown).  {\bf B.} The firing rates predicted in mean field (vertical axis) vs. the firing rates observed in the simulation of panel A via Eq.~\ref{eq:ftimeaver} (with $T=10,000$ time steps). The dashed line is the identity line. Note how the firing rates in the network span almost the entire range of admissible values, and the points are mostly located along the identity line, which confirms the good agreement between the mean field predictions and the actual firing rates.}
\label{fig:het} 
\end{figure}

\subsubsection{Heterogeneous population}

From the mean field equations \ref{eq:meanf}, we guess (see the next subsection) that they can admit a solution with different firing rates across neurons only if the external input currents (or the response functions) are different for different neurons. An example is shown in Fig.~\ref{fig:het}A for a network of excitatory neurons with a uniform distribution of external currents. The firing rates are widely distributed across neurons and are well captured by the mean field equations (panel B). This network can be interpreted as a collection of neurons with different characteristics (e.g., by choosing $\theta_i=-I_{i,ext}$ and setting all external currents to zero), and therefore it is a model of a heterogeneous network. 

There is another meaning in which a network can be considered heterogeneous, i.e., in the presence of random connectivity. This will be considered in Sec.~\ref{sec:randconn}.

\subsubsection{Homogeneous population} \label{sec:hom}

\begin{figure}[t]
\centering
\includegraphics[scale=0.4]{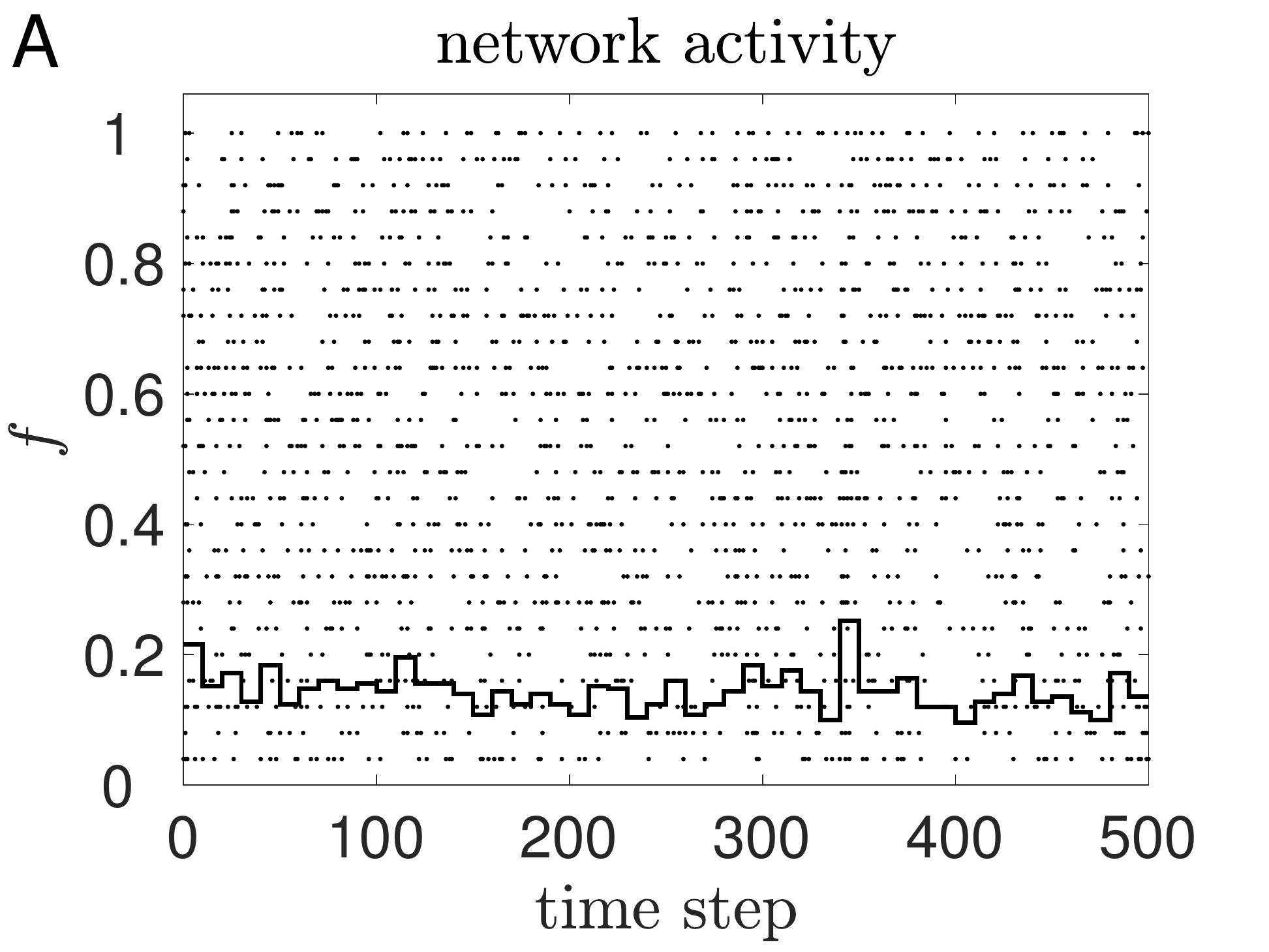}
\includegraphics[scale=0.4]{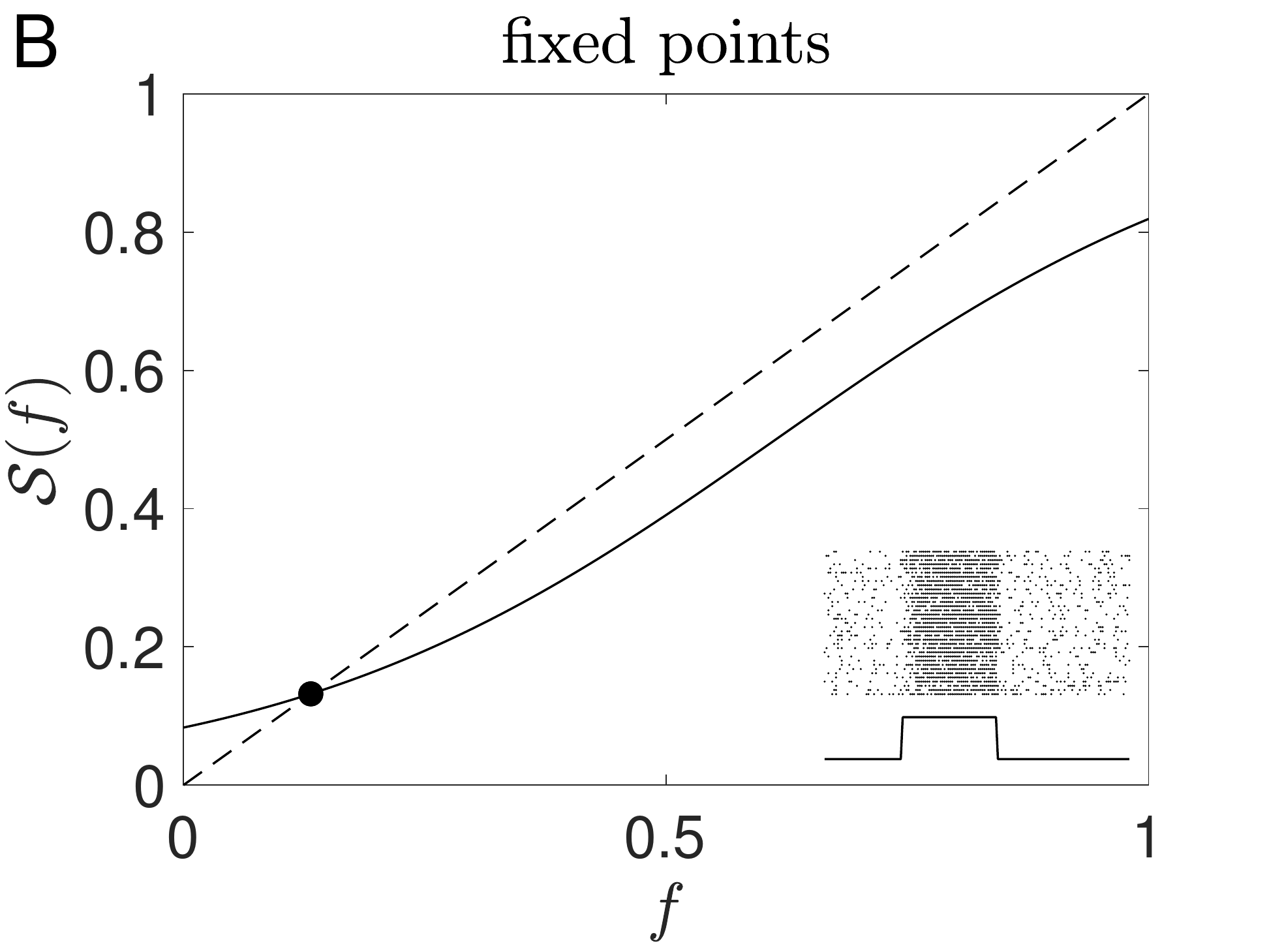}
\caption{Homogeneous network with mean field prediction. {\bf A.} Raster plot from the same network of Fig.~\ref{fig:logist}B except $N=100$ and $I_{ext}=-0.6$ for all neurons. All neurons have the same firing rate. The thick curve is the ensemble average (see appendix~\ref{app:EA}) with a fixed bin size of 10 time steps (average firing rate across time bins is $0.13$, or $13$ spikes/s in suitable units). {\bf B.} The firing rate of the neurons in A can be obtained from the graphical solution of the mean field equation~\ref{eq:meanfieldhom} with $g=1$. The dashed line represents the equation $y=f$, while the full line is $y=\mathcal S(f)$. The intersection point of these two lines (circle) gives the fixed point $f^*$ (here $f^*=0.13$, in agreement with the the firing rates observed in panel A). Since the slope of $\mathcal S(f)$ is $<1$ at the fixed point, the activity of the network is stable at this point. {\em Inset:} network's activity (top) in response to an input perturbation (bottom) shows that the fixed point is stable.}
\label{fig:hom}      
\end{figure}

If the neurons are identical and receive identical external current, then the mean field equations \ref{eq:meanf} have an evident symmetry: for large $N$, the input current will be the same for all neurons and all neurons in the network will have the same firing rate, as shown in Fig.~\ref{fig:hom}A. 

%

Under the hypothesis of equal firing rates across neurons, the mean input current becomes
\be \label{eq:Ihom}
\mu_i = J \sum_{j \neq i}^N f_j +I_{ext} = (N-1)Jf+I_{ext} \approx NJf + I_{ext},
\ee
and is the same for all neurons. Note that we have used the symbol $\mu_i$ for the mean input current. This is a customary notation and will be used extensively later on.

By using \ref{eq:Ihom}, the mean field equations \ref{eq:meanf} become $N$ copies of the scalar equation 
\be \label{eq:meanfield0}
f = \mathcal S(NJf + I_{ext}).
\ee
The graphical solution of this equation is shown in Fig.~\ref{fig:hom}B for $J=1/N$ and is $f^*=0.13$, in agreement with the firing rate observed in the simulation of Fig.~\ref{fig:hom}A. Note how the activity of a single neuron (and hence the single-neuron response function) is sufficient in this case to describe the activity of the whole population. 

\begin{remark} -- {\em The scaling of $J$ with $N$ is motivated by the fact that the input current is proportional to $N$; as the size of the population increases (as required by the mean field approximation), the input will saturate the activity of all the neurons. Instead, by taking $J = g/N$, where $g$ is a constant, Eq.~\ref{eq:meanfield0} reads
\be \label{eq:meanfieldhom}
f = \mathcal S(g f + I_{ext}),
\ee
an equation in which $N$ has disappeared -- hence valid in the infinite network. Synaptic weight scaling will be considered in more detail in Sec.~\ref{sec:scaling}.}
\end{remark}

\begin{remark} -- {\em The approach used in this section illustrates a typical reasoning of mean field theories: one lays out hypotheses that are intuitive consequences of the mean field assumptions (Gaussian current, uniform firing rates); one then derives and solves the equations; and finally checks, {\em a posteriori} and self-consistently, that the hypotheses were correct. }
\end{remark}

\subsubsection{Stability of the fixed points} \label{sec:stability}

The mean field equations also tells us about the stability of the fixed point, at least with respect to the dynamics Eqs.~\ref{eq:meanIeq}. For the model of Eq.~\ref{eq:meanfieldhom}, the dynamics is the same for all neurons and reads
\be \label{eq:meanIeq'}
\tau_I \dot I = -I + g \mathcal S(I) + I_{ext}.
\ee
A fixed point $f^*$ of this model is stable if the slope of the transfer function at $f^*$ is smaller than 1,
\be \label{eq:stability}
\left . {\partial \mathcal S \over \partial f} \right |_{f^*}  < 1,
\ee
while it is unstable if this slope is larger than one. This is a standard result of linear stability analysis, and can be understood as follows. The fixed point $x^*$ of the system $\dot x = -x + \Phi(x)$ is obtained for $\Phi(x^*)=x^*$. For $x>x^*$ but very close to $x^*$, stability requires $\dot x = \Phi(x)-x<0$ (so that $x$ will decrease back to $x^*$), which is true if $\Phi(x)$ lies below $x$. In turn, this is true if the slope of $\Phi(x)$ at the fixed point is smaller than the slope of $y=x$, i.e., for $\Phi'(x^*)<1$. One can similarly work out the other cases. Using the fact that $I^* = g f^* + I_{ext}$ together with the chain rule of derivation, we obtain Eq.~\ref{eq:stability}.

Although Eq.~\ref{eq:meanIeq'} is not the real dynamics of the network, it captures the stability of its fixed points -- as long as the network is large enough. This is shown in the inset of Fig.~\ref{fig:hom}B, where the activity returns to $f^*$ after the removal of a rather strong perturbation. Note that we only have $N=100$ in this example.

\subsubsection{Bistability} \label{sec:bist}

\begin{figure}[t]
\centering
\includegraphics[scale=.34]{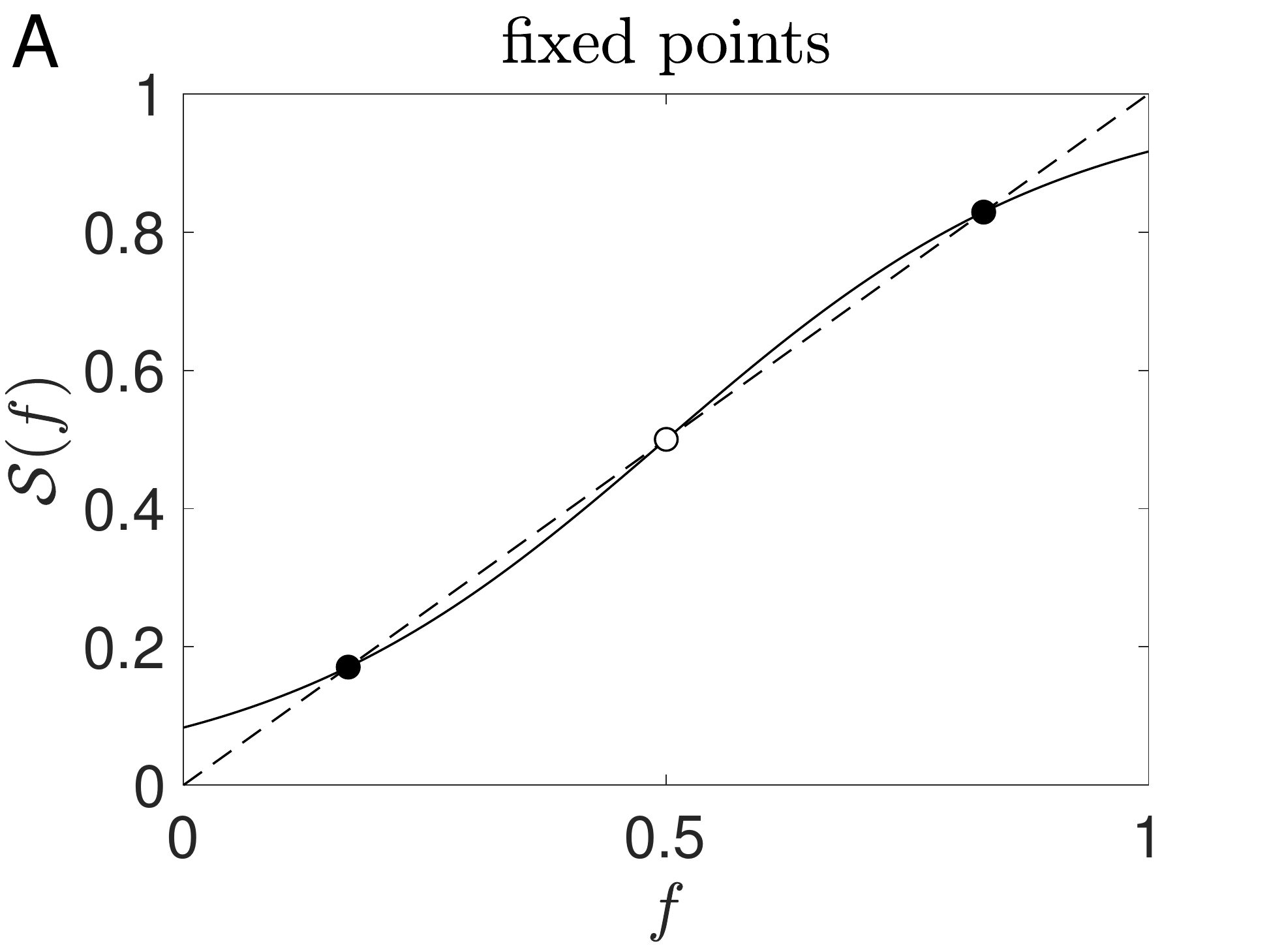} \hspace{-0.5cm} 
\includegraphics[scale=.34]{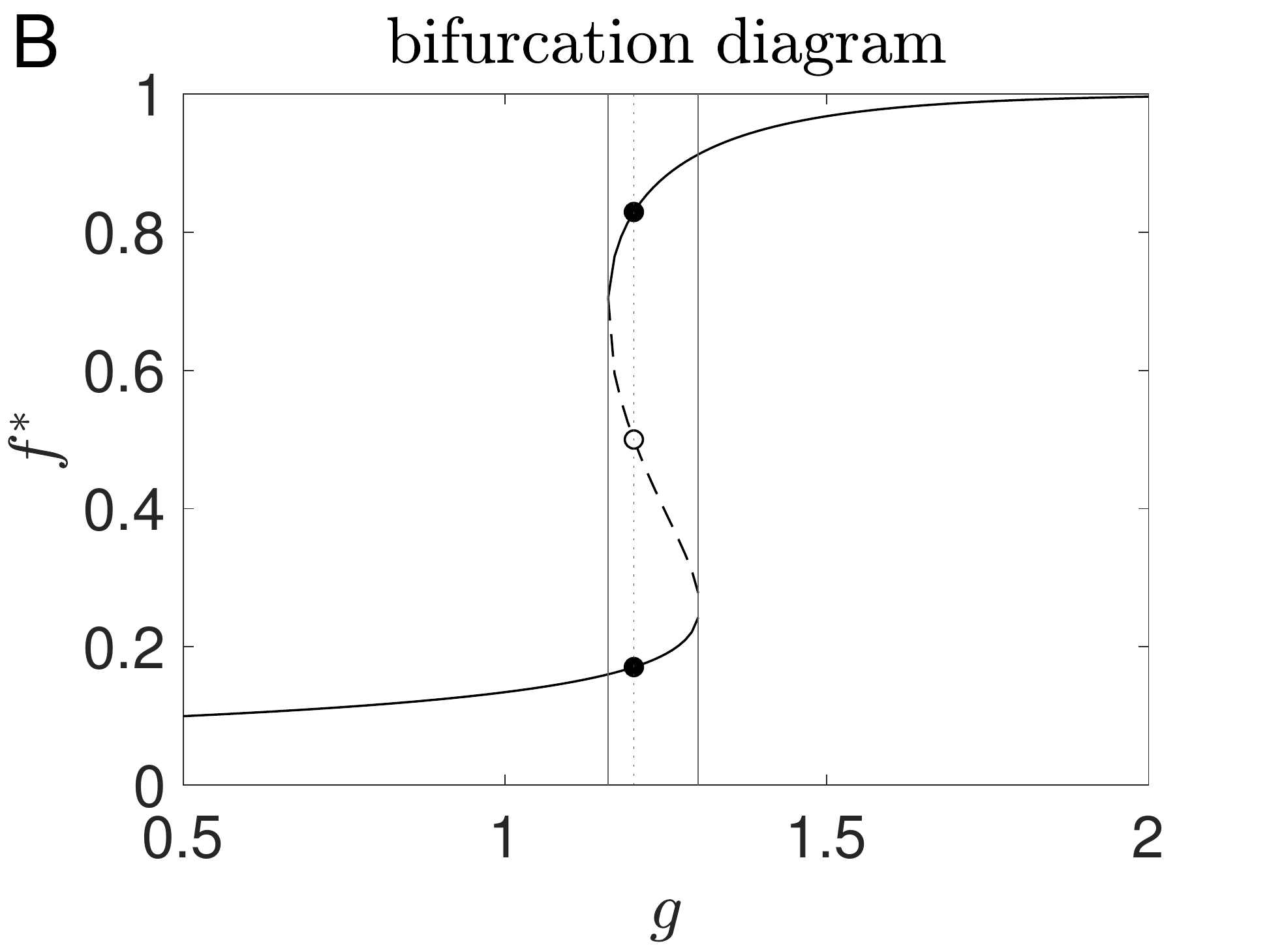}\\
\includegraphics[scale=.34]{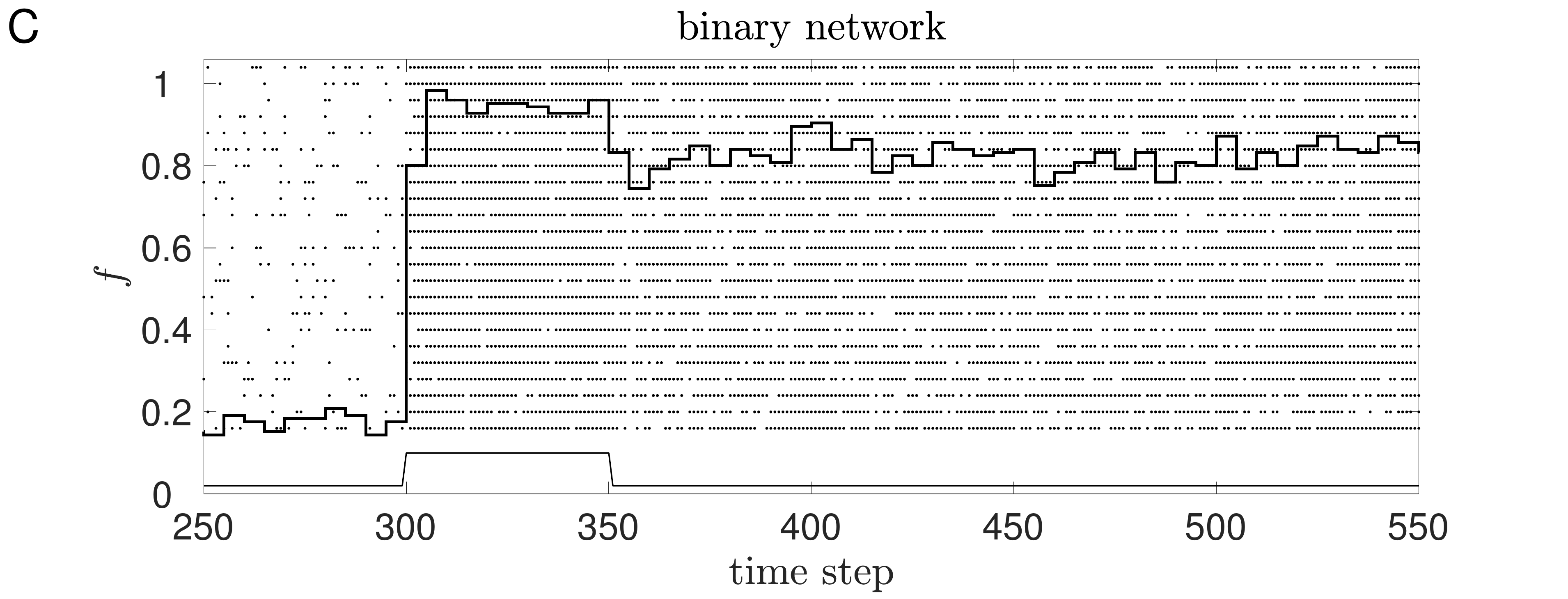} 
\caption{\small Bistability in the network of Fig.~\ref{fig:hom}. {\bf A.}  Graphical solution of the mean field equations \ref{eq:meanfieldhom} for $g=1.2$. There are now three intersection points, two of which stable (black circles, firing rates $0.17$ and $0.83$, respectively). {\bf B.} Bifurcation diagram of the network of panel~A. The plot shows the fixed points as $g$ is varied. The interval $1.16<g<1.3$ (vertical full lines) is the bistability interval with three fixed points: one unstable (on the dashed branch) and two stable. Vertical dotted line corresponds to $g=1.2$ used in panel~A. {\bf C.} Raster plot and ensemble average for the network in panel~A with $N=1,000$ (same keys as in Fig.~\ref{fig:hom}A). The activity is initially at the lower fixed point and, after a transient stimulation (shown at the bottom), enters the higher fixed point. See the text for details.}
\label{fig:bist}      
\end{figure}

For a suitable choice of parameters, the single homogeneous population of Fig.~\ref{fig:hom} can be bistable, in the sense that it can have two stable points of activity: one at low firing rate and one at high firing rate. This is shown in Fig.~\ref{fig:bist}. Note that, given $\mathcal S(x)$, the mean field equations \ref{eq:meanfieldhom} depend only on $g$ and $I_{ext}$. As $g$ is increased, the shape of $\mathcal S(f)$ will change. For $g=1.2$ there are three intersection points. Based on Eq.~\ref{eq:stability}, the middle point (white circle) is unstable while the other two are stable (black circles). This means that the network can be found in one of two stable activity regimes: one at low firing rate and one at high firing rate. 

Bistability is an important property that has been used to model perception \cite{mrr07}, memory \cite{ab97}, and decision making \cite{w02}, and therefore it is of interest to establish under what conditions a neural circuit can be bistable. This is done with the aid of a {\em bifurcation diagram}, which plots the fixed points as a function of the mean synaptic weight, as shown in Fig.~\ref{fig:bist}B. From the diagram we see that the network is bistable for $1.16<g<1.3$, whereas outside this interval the network is monostable (there is only one fixed point). Inside the bistable region, a vertical line will intersect the diagram at three points, two stable and one unstable (located on the dashed branch). These points correspond to intersection points in the related plot of panel~A. The values $g=1.16$ and $g=1.3$ are the {\em critical points}, since as they are crossed by $g$, a qualitative different behavior emerges. 

Fig.~\ref{fig:bist}C illustrates the bistable network as a model of short-term memory \cite{mc88,fbg89,a95,Miller:1996iq}. Let's assume that our network contains neurons that respond to a particular sensory stimulus (such as a visual image). In the absence of the stimulus, the network is in the lower fixed point. At time $300$, an input current mimicking the presence of the sensory input is turned on, causing the activity to rise. After the stimulus is removed, the activity settles on the higher fixed point. Since the activity at the higher fixed point persists after the removal of the stimulus, it may be interpreted as an internal representation of the stimulus. In this example with a single population one can only accommodate one memory, however this restriction can be avoided by partitioning the network into subpopulations of neurons \cite{ab97}. This model is discussed next.

%

\subsection{Clustered networks} \label{sec:clusters}

The homogeneous network is the basis for an important generalization, one in which the network in partitioned in $M$ homogeneous subpopulations, or clusters. We now consider this case. Given populations $\alpha$ and $\beta$ with {\em mean} synaptic weights $J_{\alpha \beta}$ for all $i \in \alpha$ and $j \in \beta$, proceeding as done in section~\ref{sec:randomJ} we have
\begin{equation} \label{eq:meanI'}
\langle \langle I_i \rangle \rangle = \sum_{j \neq i} \langle \langle J_{ij} x_j  \rangle \rangle = \sum_{j \neq i} [ J_{ij} ] \langle x_j \rangle = \sum_{\beta=1}^M N_{\beta} J_{\alpha \beta} f_{\beta} ,
\end{equation}
where $N_{\beta}$ is the number of neurons in cluster $\beta$ (not to be confused with the parameter of the logistic function) and $f_{\beta}$ is the population-average of the neuronal firing rates in cluster $\beta$. Note that \ref{eq:meanI'} is the same for all neurons in cluster $\alpha$. The mean firing rate of any neuron in population $\alpha$ is therefore given by the mean field equations:
\be \label{eq:mfalpha}
\boxed{f_{\alpha} = \mathcal S \left (\sum_{\beta=1}^M N_{\beta} J_{\alpha \beta} f_{\beta} + I_{{\alpha},ext} \right), \quad \alpha =1, ..., M,}
\ee
where we have assumed all neurons of the same population receive the same external current. Note that now determining the fixed points and their stability requires a generalization of the analysis of Sec.~\ref{sec:stability} \cite{ma99,Mazzucato2016-hl}.

One special case of clustered network is the excitatory-inhibitory recurrent network. In this case we have two populations, one having excitatory ($E$) neurons and one having inhibitory ($I$) neurons, and 4 types of mean synaptic weights $J_{\alpha \beta}$: $J_{EE}$, $J_{II}$, $J_{EI}$ and $J_{IE}$, where e.g. $J_{EI}$ are the mean synaptic weights from inhibitory to excitatory neurons. Note that this model respects Dale's law, stating that neurons can be excitatory or inhibitory, but not both.

Clustered networks are much studied, especially in the context of integrate-and-fire neurons. Since the fluctuations of the neural activity play an essential role in clustered networks, we first show how to incorporate these ingredients in the theory, and defer a discussion of clustered networks to the end of Sec.~\ref{sec:lif}.

\section{Extensions} \label{sec:extensions}

In this section we consider two very important extensions of the theory, the incorporation of the variability of the input current generated by the network itself, and the random connectivity of the neurons. We start from the former. 
  
\subsection{The impact of the input variance on the mean firing rates} \label{sec:variance}

In deriving our mean field approximation, we have replaced the current with its mean input. The input is the sum of many contributions and therefore, by the central limit theorem, converges to a Gaussian random variable in the thermodynamic limit $N \to \infty$. A Gaussian distribution is characterized by its mean and variance. Hence, by incorporating the impact of the variance of $I_i$ into our model, we can go beyond mean field and provide a more accurate description of the network's behavior. 

\begin{figure}[t]
\centering
\includegraphics[scale=.4]{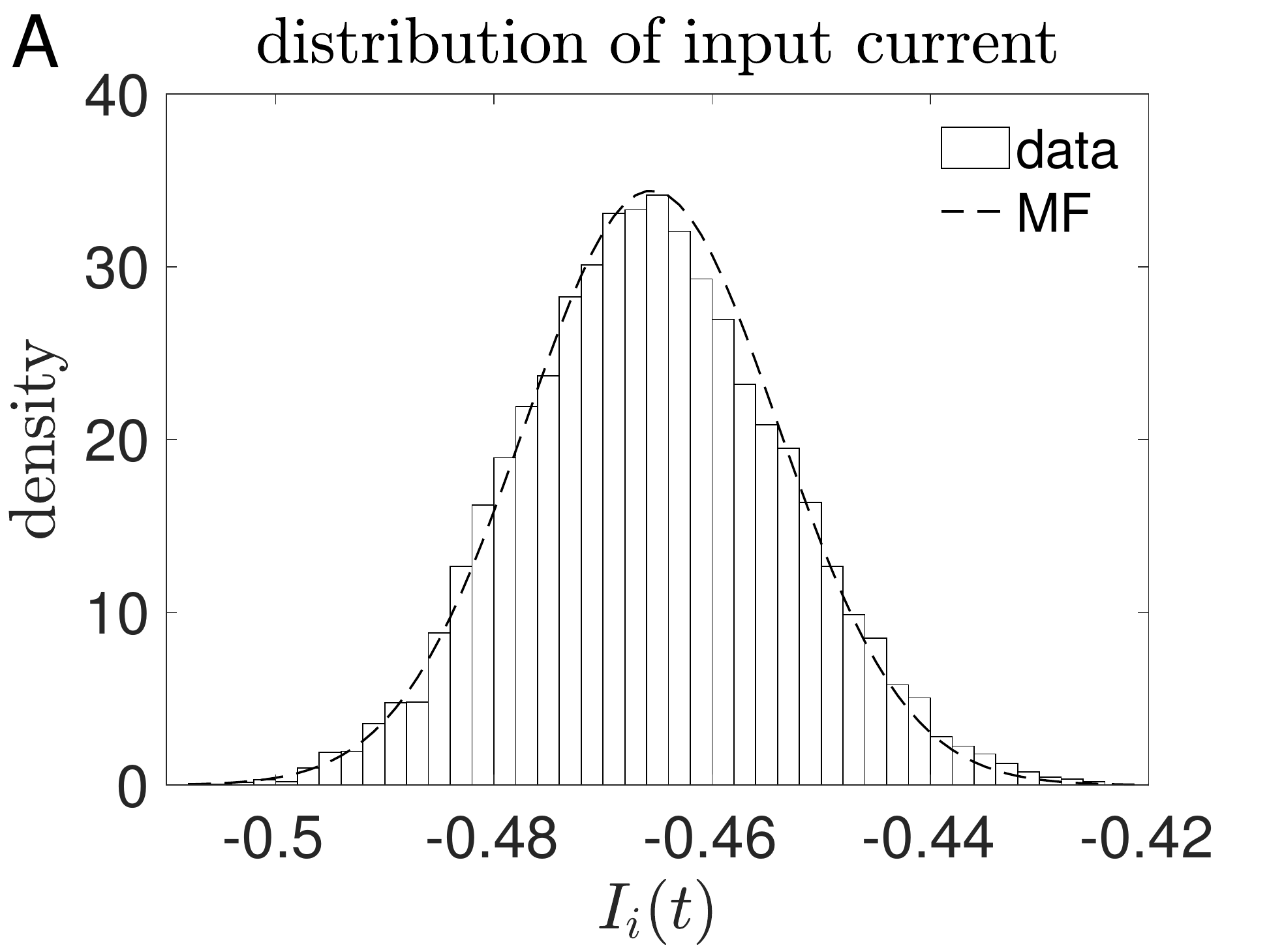}
\includegraphics[scale=.4]{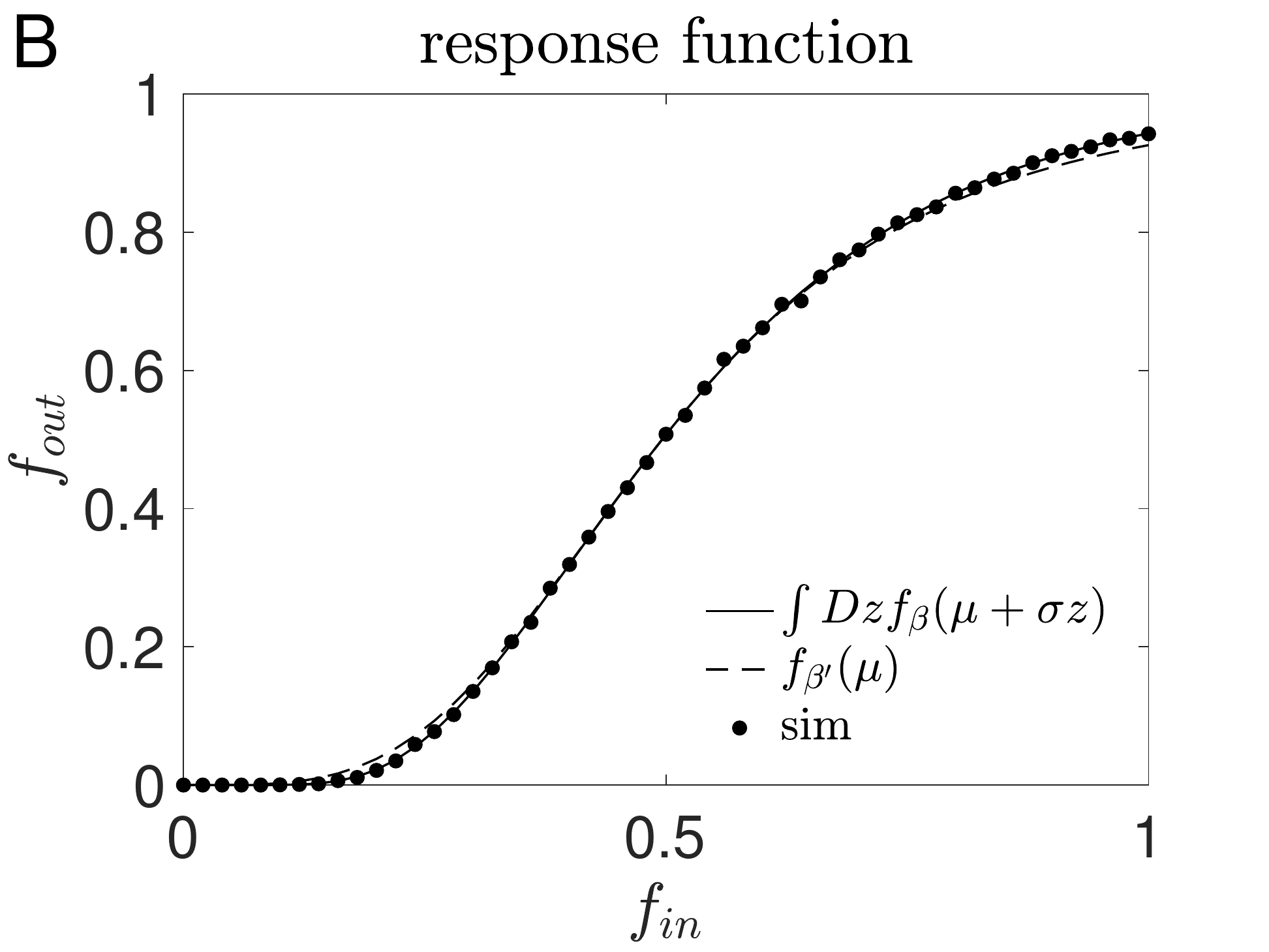}
\caption{\small {\bf A.} Distribution of input currents across time for the network of Fig.~\ref{fig:hom} at the fixed point. To avoid variations in current across neurons, equal weights equal to $g/N$ were used (with $N=1000$). The distribution is described well by a Gaussian distribution with mean and variance predicted by mean field theory Eqs.~\ref{eq:mu-sigma-bin} (dashed). {\bf B.} Response function of the noise-driven logistic neuron: comparison of Eq.~\ref{eq:fifull} (full line) and its approximation Eq.~\ref{eq:filogist} (dashed) with simulations of the logistic neuron driven by Gaussian input current (dots). The output firing rates are functions of the input firing rates through $\mu_i(f_{in}), \sigma_i(f_{in})$ given by Eqs.~\ref{eq:mu-sigma-bin}.}
\label{fig:curr} 
\end{figure}

We can deduce heuristically the effect of Gaussian fluctuations arguing as follows. We replace the input current with its Gaussian approximation valid in the large $N$ limit:
\begin{equation} \label{eq:I2}
I_i(t) = \sum_{j \neq i}^N J_{ij} x_j(t) + I_{i,ext} \approx \langle I_i \rangle + \eta_i(t) \equiv \mu_i + \sigma_i z(t),
\end{equation}
where $z(t) \sim \mathcal N(0,1)$ is a standard Gaussian variable and $\mu_i, \sigma_i^2$ are the mean and variance of $I_i$, respectively. Fig.~\ref{fig:curr}A shows that the temporal fluctuations of the input current are indeed well described by a Gaussian distribution.

%

For convenience, the constant term $I_{i,ext}$ has been included into the mean $\mu_i$. The firing rate, for a given value of $z(t) = z$, is given by Eq.~\ref{eq:fix}
\be
f_i(z) = { 1 \over 1 + e^{-2\beta (\mu_i + \sigma_i z - \theta)} },
\ee
where $\beta$ is constant. Mean field amounts to setting $z=0$. Now we can relax this hypothesis and compute the average firing rate Eq.~\ref{eq:fgauss} by adding up all contributions $\mu_i + \sigma_i z$, each weighted by the probability of $z$:
\be \label{eq:fifull}
f_i(\mu_i, \sigma_i) = \int_{-\infty}^{+\infty} {dz \over \sqrt{2 \pi} } e^{-{z^2 \over 2}} { 1 \over 1 + e^{-2\beta (\mu_i + \sigma_i z - \theta)} }.
\ee
This is the response function of the binary logistic neuron when the fluctuations of the input current are taken into account, as shown in Fig.~\ref{fig:curr}B. This function is closely approximated by a logistic function with a different parameter $\beta'$ (see e.g. \cite{Maragakis:2008zb} and Fig.~\ref{fig:curr}B):
\be \label{eq:filogist}
f_i(\mu_i, \sigma_i) \approx { 1 \over 1 + e^{-2\beta' (\mu_i - \theta)} }, 
\ee
where
\be \label{eq:beta'}
2 \beta'_i = \left ( {1 \over 4 \beta^2} + { \pi \sigma_i^2 \over 8 } \right )^{-1/2}.
\ee
Now there are two sources of noise: $\beta$ and $\sigma_i$, where the latter originates from the activity of the network itself. 

When $\sigma_i$ is small, this function approaches $f_i=\mathcal S(\mu_i)$, i.e. Eq.~\ref{eq:fI} evaluated at the mean current, from which we recover our mean field equation Eq.~\ref{eq:meanf0}. When $\sigma_i$ is large enough, though, it endows the network with its own source of variability due to Gaussian nature of the input current. Hence, it is no longer necessary to assume an intrinsic form of noise $\beta$, and we can allow our model to be deterministic by setting $\beta \to \infty$. In this limit one immediately gets:
\be \label{eq:filargebeta}
f_i(\mu_i, \sigma_i) \to { 1 \over 1 + e^{-\sqrt{8 \over \pi} \left ({ \mu_i - \theta \over \sigma_i} \right ) } }.
\ee

Comparison with Eq.~\ref{eq:Pspike} shows that this is our previous response function $\mathcal S(\mu_i)$ with $\beta \propto \sigma_i^{-1}$ (recall that in Eq.~\ref{eq:Pspike} we had set $\theta=0$). There are important differences, however:

\begin{itemize}

\item now the noise affecting the spike probability (hence the firing rate) is the result of the random input current rather than intrinsic noise in the spiking mechanism (which is now deterministic);

\item unlike $\beta$, $\sigma_i$ is not constant but depends on the activity of the network -- in particular, it depends on the firing rates of the other neurons, as we shall see shortly;

\item Eq.~\ref{eq:filargebeta} shows that the firing rate is a sigmoidal function of $\mu_i$, with $\sigma_i$ controlling its slope.

\end{itemize}

It turns out that this heuristic picture, including the dependence of the firing rate on $\mu_i$, $\sigma_i$ in the form $(\mu_i-\theta)/\sigma_i$, is correct also for more realistic models of spiking neurons (Sec.~\ref{sec:lif}). The reason is intuitively simple: the firing rate is determined by the distance between $\mu_i$ and $\theta$ {\em in units of $\sigma_i$}: in the presence of noise, the difference $\mu_i-\theta$, on its own, is not sufficient to determine the firing rate. 

\begin{remark} --  {\em What does, in the deterministic model where $\beta \to \infty$, make the input current behave as a stochastic variable? This has to do with the chaotic nature of the dynamics resulting from ingredients such as quenched synaptic weights, random connectivity (discussed later) and the recurrent nature of the network. More details will be given later.}
\end{remark}

\subsubsection{The moments of the input current} \label{sec:moments}

To close the self-consistency loop of the mean field equations, we need to determine the dependence of $\mu_i, \sigma_i$ on the firing rates of the presynaptic neurons. For the mean we have Eq.~\ref{eq:meanI2}, which we write here in the equivalent form:
\be
\mu_i = \sum_j [J_{ij}] f_j + I_{i,ext}.
\ee
To compute the variance we can use the formula for the variance of the product of two independent random variables applied to $J_{ij} x_j$ (in the following $\mathbb E(z)$ denotes the generic expectation of $z$, and note that $x_j^2=x_j$):
\begin{eqnarray}  \nonumber
\text{Var}(J_{ij} x_j) & = & \text{Var}(J_{ij}) \; \mathbb E(x_j^2) + \text{Var}(x_j) \; \mathbb E^2(J_{ij}) \\ \nonumber
& = & \text{Var}(J_{ij}) f_j  + f_j (1-f_j) \; \mathbb E^2(J_{ij}) \label{eq:varJxexact} \\
& = & ( \text{Var}(J_{ij}) + \mathbb E^2(J_{ij}) ) f_j - \mathbb E^2(J_{ij}) f_j^2 \\
& \approx & \mathbb E(J_{ij}^2) f_j,  \label{eq:varJx}
\end{eqnarray} 
where the approximation is valid for small $f_j$, which is a relevant case in cortex. Although we know the exact result Eq.~\ref{eq:varJxexact}, we chose to emphasize the approximate result in \ref{eq:varJx} because, as we shall see later, there is a sense in which this result is exact in networks of spiking neurons. Also, low firing rates tend to decorrelate the activities of the neurons, an assumption required to apply the central limit theorem and to add up the variances coming from the $N$ neurons of the network, which from Eq.~\ref{eq:varJx} gives
\be \label{eq:sigma_binary}
\sigma_i^2 \approx \sum_j^N [J^2_{ij}] f_j.
\ee
For the homogeneous population of Fig.~\ref{fig:curr} where the synaptic weights were set equal to $g/N$, using these formulae we obtain
\be \label{eq:mu-sigma-bin}
\mu_i = g f + I_{ext}, \quad \sigma_i^2 \approx {g^2 f \over N},
\ee
which are the same for all neurons. The Gaussian density function with these parameters predicts well the current's temporal fluctuations, as shown in Fig.~\ref{fig:curr}A. Note how in this case the variance will vanish in the thermodynamic limit due to our choice $J_{ij} \sim 1/N$ (but see Sec.~\ref{sec:scaling}).

\subsubsection{Extended mean field theory}

When taking into account the variance of the input, the self-consistent mean field equations read, in vectorial notation, as
\be \label{eq:mfsigma}
{\pmb f} = \pmb{\mathcal S} ({\pmb \mu({\pmb f}), {\pmb \sigma}({\pmb f})}).
\ee

An important example is the clustered network of Sec.~\ref{sec:clusters}. In that case, all neurons in the same cluster receive current with the same input and variance, and we obtain
\begin{equation} \label{eq:mu-sigma}
\mu_{\alpha} = \sum_{\beta=1}^M N_{\beta} [J]_{\alpha \beta} f_{\beta} + I_{\alpha,ext}, \quad \sigma^2_{\alpha} = \sum_{\beta=1}^M N_{\beta} [J^2]_{\alpha \beta} f_{\beta},
\end{equation}
where $[J]_{\alpha \beta}$ and $[J^2]_{\alpha \beta}$ are the mean and the second moment of the synaptic weights between neurons of populations $\alpha$ and $\beta$. Note that while the terms in the mean input can be positive or negative (depending on the sign of $[J]_{\alpha \beta}$, which is negative if population $\beta$ is inhibitory), the variance is the sum of positive terms. This is because we have assumed that the inputs coming from different neurons are independent, i.e., their covariances vanish.

\vsp Strictly speaking, the theory is valid in the thermodynamic limit, which in the clustered network requires some care. Approximately, however, we expect good predictions for a large enough number of neurons in each cluster. Also, it is possible to have situations in which $\sigma^2_{\alpha} \to 0$ in the limit (see e.g. Eqs.~\ref{eq:mu-sigma-bin}), and if no external fluctuations are added, the network's behavior becomes deterministic. We discuss ways to keep a finite variance for $N \to \infty$ in Sec.~\ref{sec:scaling}.

\subsection{Random connectivity} \label{sec:randconn}

So far, all neurons were connected to all other neurons in the network (with the exclusion of themselves). In real cortical circuits, however, neurons are connected to different numbers and types of other neurons. Even neglecting the heterogeneity in cell types, random connectivity can have a meaningful impact on the dynamics of the network and its stationary activity regimes. 

A simple, and widely used, model of random connectivity is to assume that any two neurons are connected by a synapse with probability $c$. Calling $c_{ij} \in \{0,1\}$ the random variable representing whether or not a synaptic connection exists from presynaptic neuron $j$ to postsynaptic neuron $i$, its mean and variance are $c$ and $c(1-c)$, respectively. To leverage our previous result Eq.~\ref{eq:varJx}, now in need of generalization, it is convenient to redefine the synaptic weight to include $c_{ij}$:
\be
J_{ij} \to c_{ij} J_{ij} \doteq \hat J_{ij}.
\ee
We further assume that $c_{ij}$ and $J_{ij}$ are independent random variables (i.e., synapses of different strength are equally likely to exist). It follows that the mean input in mean field becomes  
\be \label{eq:meanI2c}
\mu_i  = c \sum_j [J_{ij}] f_j + I_{i,ext},
\ee
while for the variance we have, from Eq.~\ref{eq:varJx} (note that $c_{ij}^2=c_{ij}$),
\begin{eqnarray}
\text{Var}(\hat J_{ij} x_j) \approx  c \; \mathbb E(J_{ij}^2) f_j,
\end{eqnarray}
and therefore
\be
\sigma_i^2  \approx c \sum_j [J^2_{ij}] f_j.
\ee
For a network with $M$ clusters, denoting with $c_{\alpha \beta}$ the mean connectivity from neurons in clusters $\beta$ to neurons in clusters $\alpha$, summing up over the neurons in each cluster we obtain the generalization of Eq.~\ref{eq:mu-sigma}: 
\begin{equation} \label{eq:c-mu-sigma}
\boxed{\mu_{\alpha} = \sum_{\beta=1}^M c_{\alpha \beta} N_{\beta} [J]_{\alpha \beta} f_{\beta} + I_{\alpha,ext}, \quad \sigma^2_{\alpha} = \sum_{\beta=1}^M c_{\alpha \beta} N_{\beta} [J^2]_{\alpha \beta} f_{\beta}.}
\end{equation}
The fixed points (and their stability) can be found with a first-order dynamics for the coupled vectors $\{\mu_{\alpha}, \sigma^2_{\alpha}\}$, which generalizes the methods of Sec.~\ref{sec:stability}, see e.g. \cite{ma99,Mazzucato2016-hl}.

\begin{remark} -- {\em In some models the input current to population $\alpha$ comes from $N_{ext}$ Poisson spike trains with rate $f_{ext}$, connectivity $c_{\alpha,ext}$ and synapses $J_{\alpha,ext}$, in which case the external current has both a mean and a variance,
\be \label{eq:muext}
\mu_{\alpha,ext}=c_{\alpha,ext} N_{\alpha,ext} [J]_{\alpha,ext} f_{ext}, \quad \sigma^2_{\alpha,ext}=c_{\alpha,ext} N_{\alpha,ext} [J^2]_{\alpha,ext} f_{ext},
\ee
which enter the right hand sides of Eqs.~\ref{eq:c-mu-sigma}.}
\end{remark}
In the next section we introduce clustered networks of spiking neurons in continuous time, and we'll see that the mean field equations are given, also in that case, by Eqs.~\ref{eq:mfsigma} and~\ref{eq:c-mu-sigma}. The only difference will be in the sigmoidal response function $\mathcal S$.

In the example considered in this section, connections among neurons are made randomly and independently with a fixed probability, a structure sometimes called {\em Erd\"os-R\'enyi} connectivity. However, mean field theory can also be developed in networks with more complex connectivity structures (see e.g. \cite{Nykamp:2017jy}).

\section{Mean field theory for networks of integrate-and-fire neurons} \label{sec:lif}

The theory developed so far can be applied to networks of integrate-and-fire neurons. This is a more relevant case because of its greater biological significance and the possibility for the theory to be directly tested in experiment.

\subsection{Leaky integrate-and-fire neuron} 

For concreteness, we shall develop the theory for networks of leaky integrate-and-fire (LIF) neurons. LIF neurons are characterized by their membrane potential $V(t)$ at time $t$ according to the standard model
\begin{equation} \label{eq:V}
{dV_i \over dt} = -{V_i-V_{L} \over \tau} + \sum_{j\neq i}^N J_{ij} \sum_k \delta(t-t_k^j) + I_{i,ext}.
\end{equation}
Here, $V_{L}$ is the resting potential, $\tau$ is the membrane time constant, $J_{ij}$ are the synaptic weights in voltage units, $\delta(t)$ is Dirac's delta function, and $t_k^j$ is the time of the $k^{th}$ spike emitted by presynaptic neuron $j$. The two rightmost terms represent the input current: the {\em synaptic} and external current, respectively. Note that both terms are in units of voltage/time; to obtain these terms in units of current, one should divide them by $C_m$, the membrane capacitance. To simplify the formule, here we assume $C_m=1$ and keep the input current in units of voltage/time. When the inputs contain excitatory and inhibitory spike trains, this model goes also under the name of Stein model \cite{s65}. Since this model lacks the non-linear conductances responsible for action potential generation, we complement it with {\em  boundary conditions} on $V$ to mimic the emission of a spike. Specifically: When $V$ hits a threshold $\theta$, a spike is said to be emitted and $V$ is immediately reset to a value $V_r$, where it is clamped for a refractory period $\tau_r$. After a time $\tau_r$, the dynamics Eq.~\ref{eq:V} resumes. 

\begin{figure}[t]
\centering
\includegraphics[scale=.4]{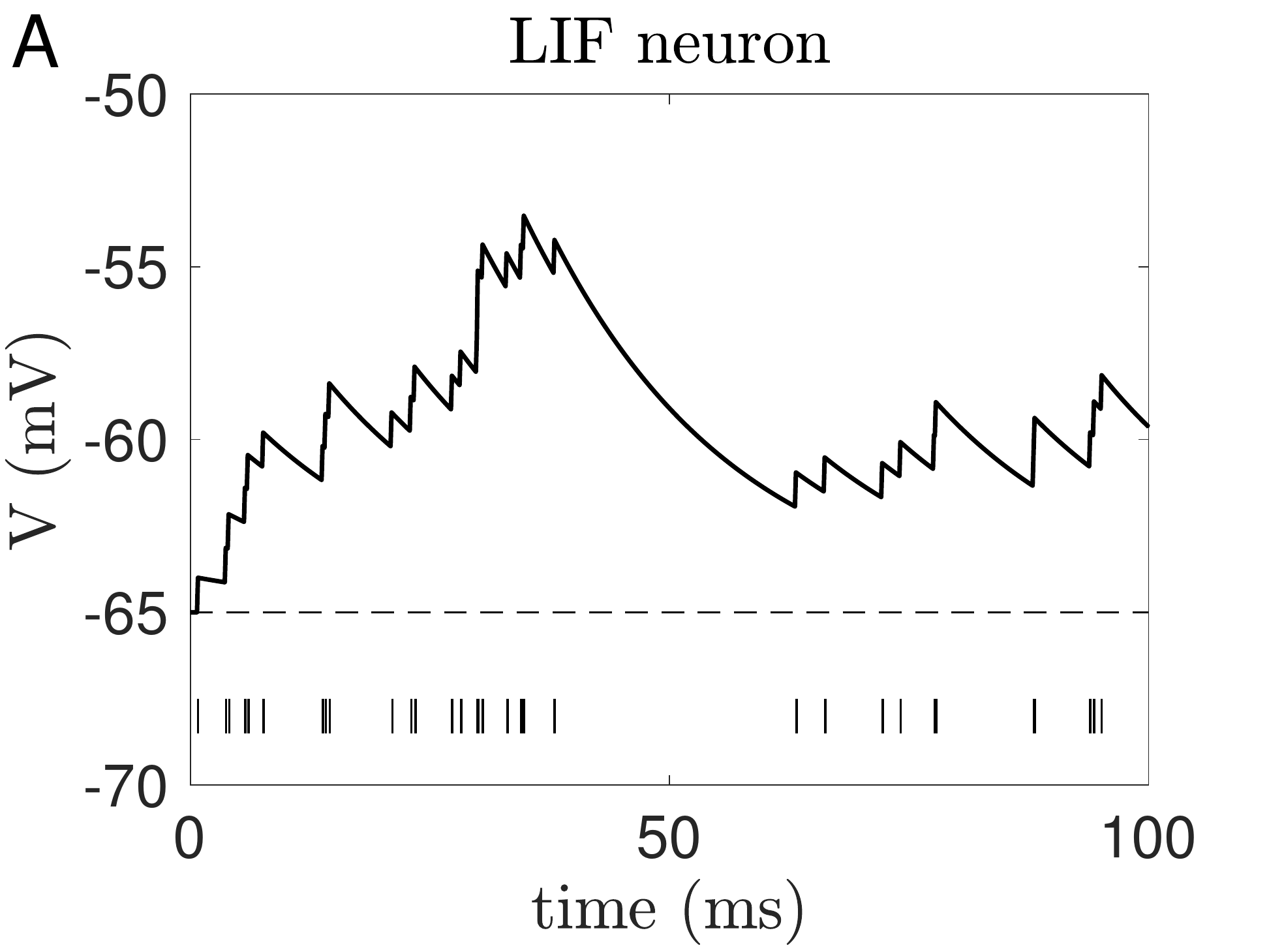}
\includegraphics[scale=.4]{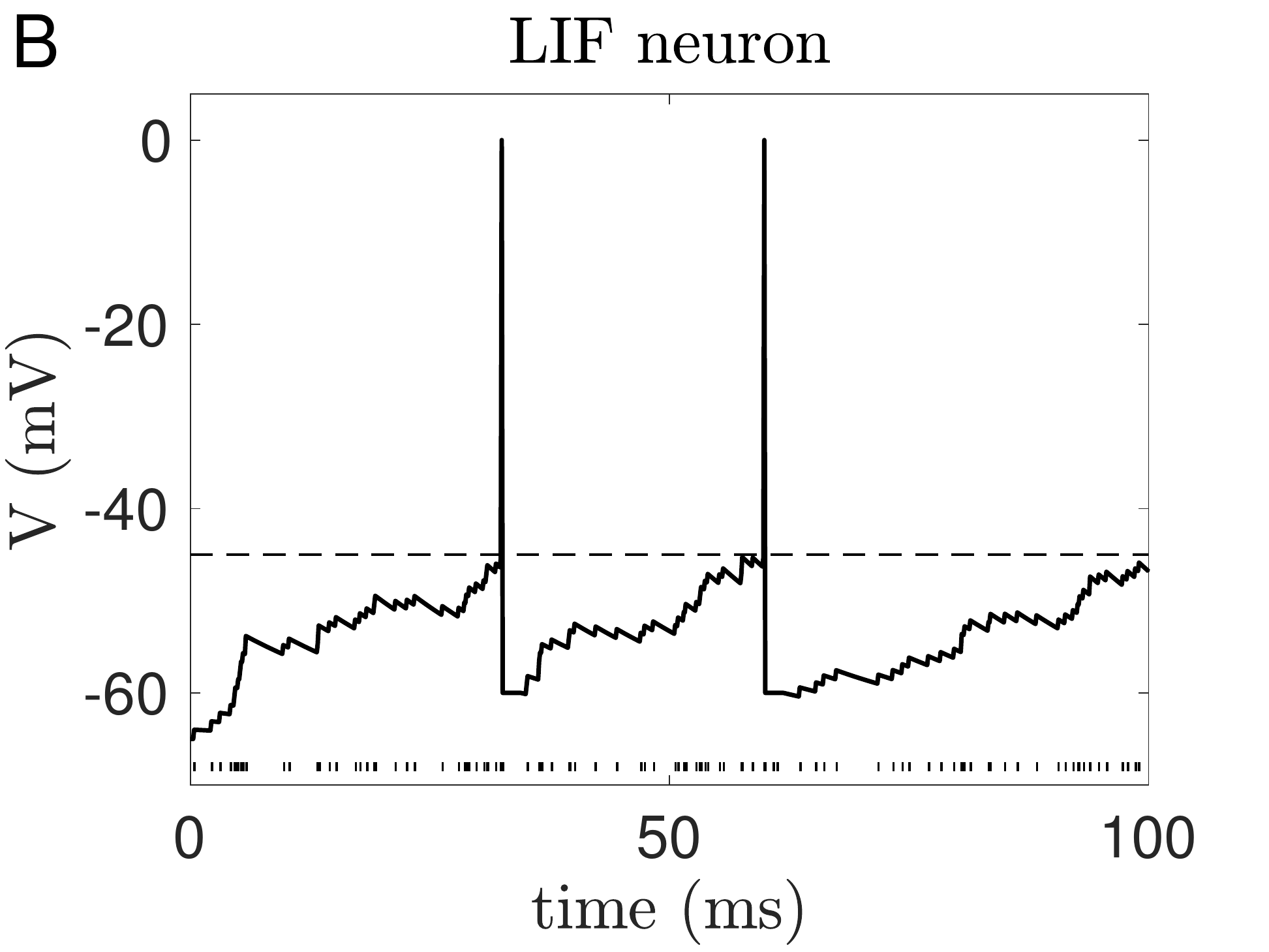}
\caption{\small Membrane potential of the LIF neuron, Eq.~\ref{eq:V}, driven by one excitatory Poisson spike train with firing rate $300$~Hz for sub-threshold input (panel A) and $1200$~Hz for supra-threshold input (panel B). The input spike train is shown at the bottom of each panel (vertical ticks). Neuron parameters: $V_L=-65$~mV (dashed line in A), $\theta=-45$~mV (dashed line in B), $V_r=-60$~mV, $\tau_r=2$~ms, $\tau=20$~ms, $J=1$~mV. The external current was set to zero.}
\label{fig:V}
\end{figure}

This behavior is illustrated Fig.~\ref{fig:V} for the case of a single excitatory input spike train and synaptic weight $J$. As shown in the left panel, $V$ jumps by $J$ upon arrival of a presynaptic spike, and decays exponentially in between spikes, in keeping with the solution to Eq.~\ref{eq:V}:
\be \label{eq:Vsol}
V_i =  V_i(0) e^{-t /\tau}  + V_{i,L}^* (1-e^{-t /\tau}) + \sum_{jk} J_{ij} e^{-(t-t_k^j)/\tau} \Theta(t-t_k^j),
\ee 
where $V_{i,L}^* \doteq V_L+\tau I_{i,ext}$ is a constant term that represents the new equilibrium value of the membrane potential in the presence of a constant external current. Fig.~\ref{fig:V}B shows the emission of spikes (followed by a reset) when $V$ hits the threshold $\theta=-45$~mV. 

\subsection{The moments of the free membrane potential}

Analogously to the situation with the binary neuron, we need to determine the response function of this model neuron, i.e., its firing rate as a function of the input current. After reabsorbing $I_{i,ext}$ into $V_{i,L}^*$, the input current is given by the synaptic input current, i.e. (see Eq.~\ref{eq:V})
\be \label{eq:Ilif}
I_i(t) = \sum_{j\neq i}^N J_{ij} \sum_k \delta(t-t_k^j).
\ee
However, the neuron emits a spike when $V_i$, not $I_i$, exceeds the threshold. Assuming a stationary input, after a transient $V_i$ reaches the steady state (from Eq.~\ref{eq:Vsol}): 
\be  \label{eq:Vsolss}
V_i(t) =  V_{i,L}^*  + \sum_{jk} J_{ij} e^{-(t-t_k^j)/\tau} \Theta(t-t_k^j).
\ee

We are therefore interested in characterizing this term. 
 
Just as before, $V_i$ is the sum of contributions coming from many neurons. Assuming independent or, at most, weakly correlated neurons, $V_i$ follows approximately a Gaussian distribution with mean $\mu_i$ and variance $\sigma^2_i$. Let's indicate with $J_E$ the excitatory weights and with $J_I$ the inhibitory ones. Moreover, we assume that the inputs $\sum_k \delta(t-t_k^j)$ are independent Poisson spike trains with mean $f_E$ and $f_I$, respectively. Then (see appendix~\ref{app:mu-sigma} for details):
\be \label{eq:mu-sigma-2}
\mu_i = V_{i,L}^* + N_E [J_E] f_E \tau - N_I [J_I] f_I \tau, \quad  \sigma_i^2 = { 1 \over 2 } N_E [J_E^2] f_E  \tau + { 1 \over 2 } N_I [J_I^2] f_I \tau.
\ee
Note that Eqs.~\ref{eq:mu-sigma-2} are valid for the {\em free} membrane potential, i.e., in the absence of output spikes. Nevertheless, $\mu_i$ and $\sigma_i$ also determine the firing rate of the neuron, as we show next.

\subsection{The response function of the LIF neuron}

\begin{figure} 
\centering
\includegraphics[scale=0.4]{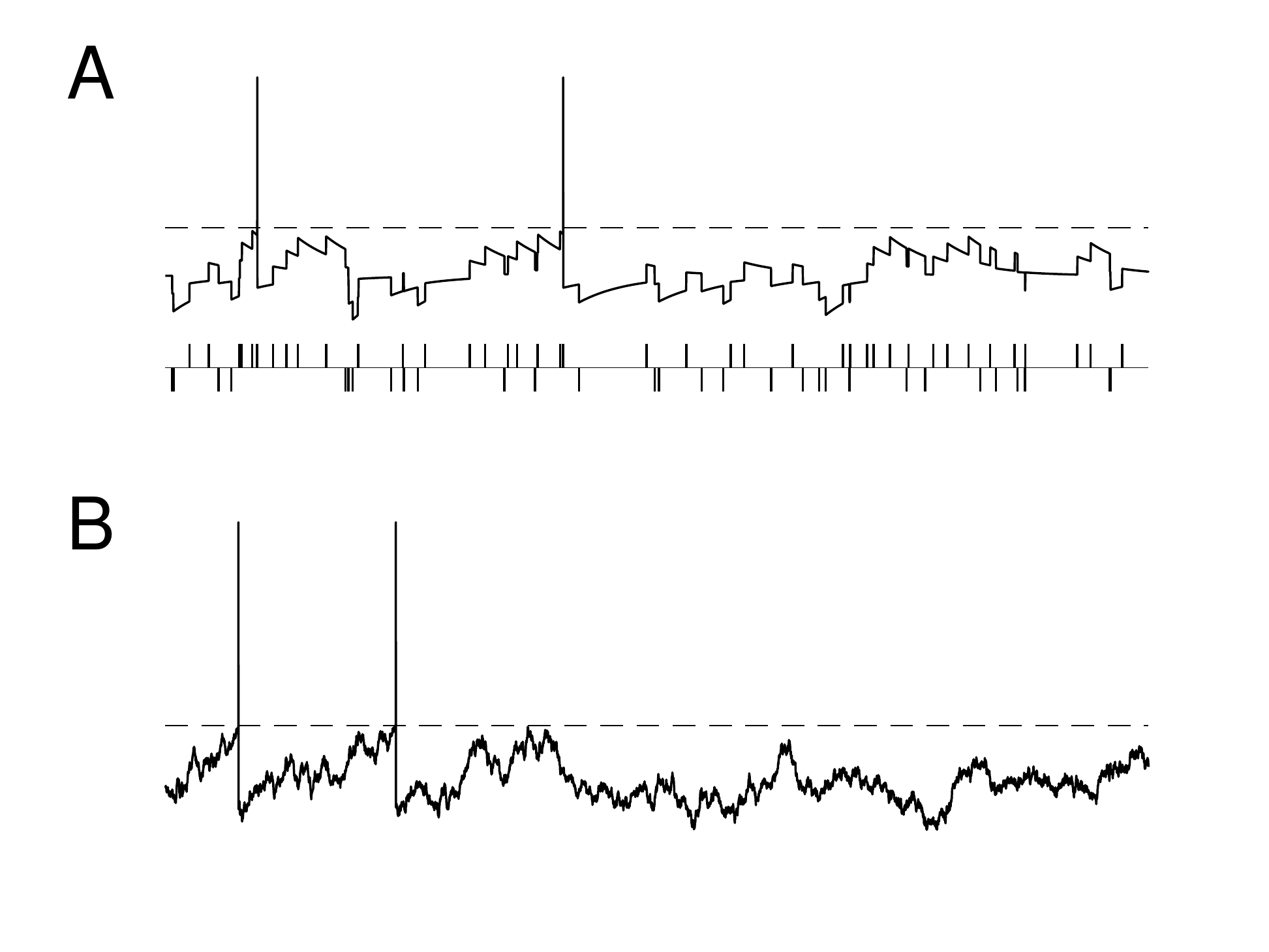}
\includegraphics[scale=0.4]{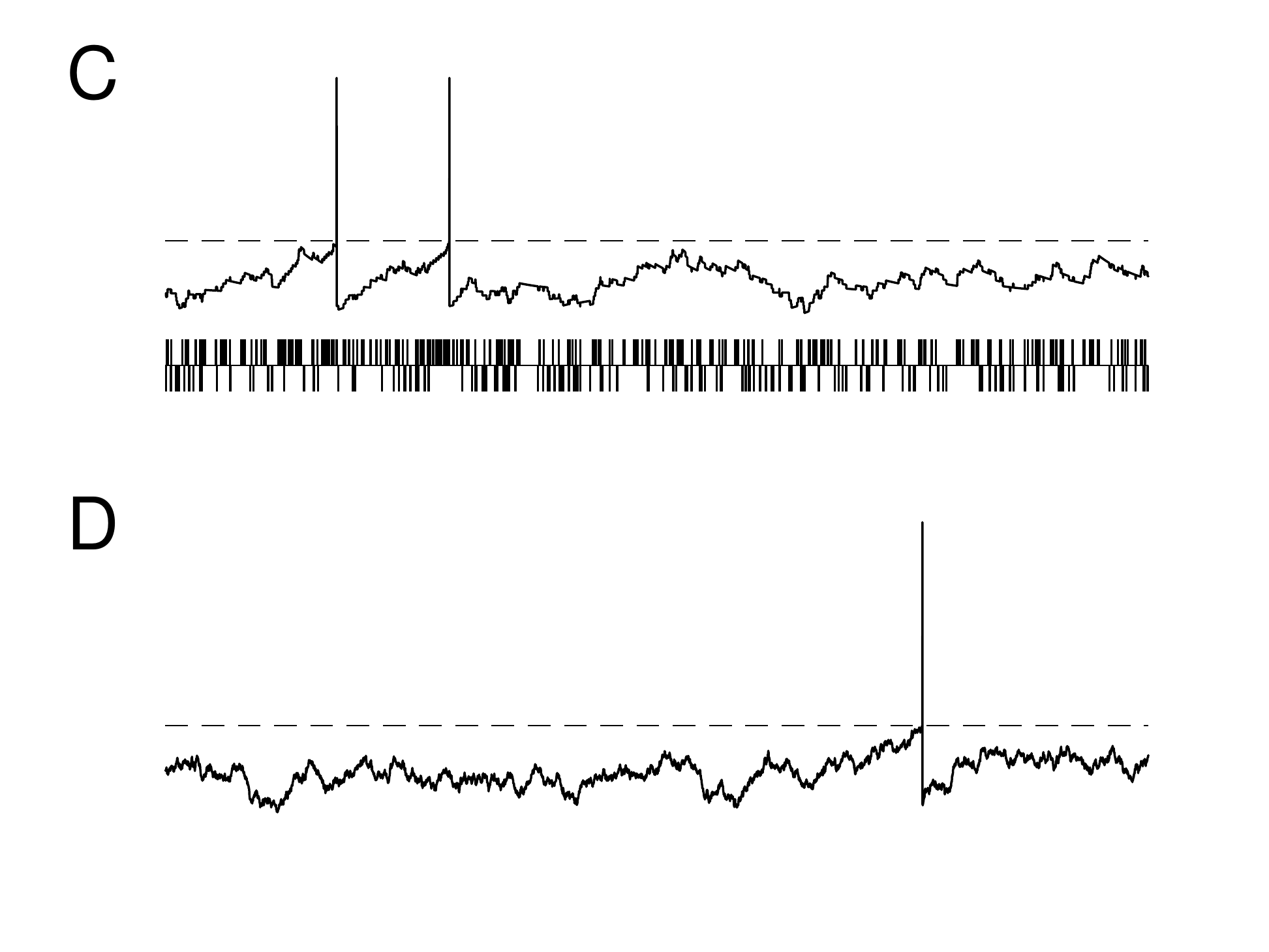}
\vspace{-0.8cm}
\caption[]{\small Diffusion approximation for the LIF neuron. {\bf A:} LIF neuron response to two spike trains (shown below the membrane potential trace), one excitatory (upward tickmarks) and one inhibitory (downward tickmarks). {\bf B:} Diffusion approximation to A: same neuron driven by fluctuating Gaussian input current with the same mean and variance as the Poisson input in A. Note that although the membrane potential and the spike times differ in the two cases, the firing rates of the output spike trains match. {\bf C:} Same as A for smaller $J$ but larger input firing rates. The membrane potential looks smoother than in A and already rather similar to its diffusion approximation shown in D. {\bf D:} Diffusion approximation to C.}
\label{fig:diff}
\end{figure}

The response function of the LIF neuron is difficult to compute despite the simplicity of the model. Fortunately, a closed formula is known under the so-called {\em diffusion approximation}, an approximation valid when 
\begin{itemize}
\item[i)] the number of presynaptic inputs is large but each synaptic input contributes a very small perturbation to the membrane potential; and 
\item[ii)] the values of the input current in successive time bins are independent (this is true if e.g. the input current is the sum of independent Poisson spike trains). 
\end{itemize}
The diffusion approximation is pictorially illustrated in Fig.~\ref{fig:diff}. 

\begin{figure}[t]
\centering
\includegraphics[scale=.45]{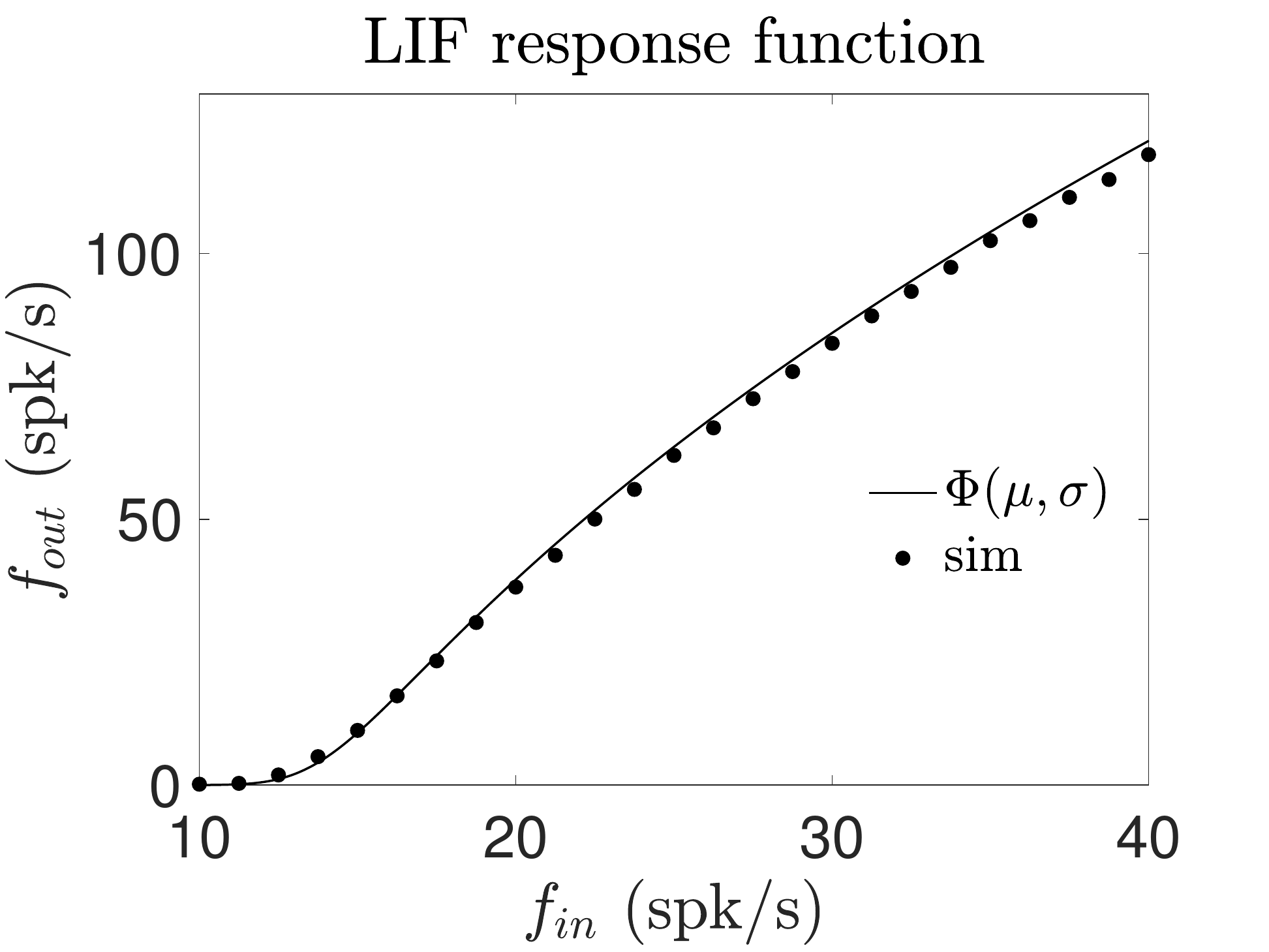}
\caption{\small Response function of the LIF neuron driven by synaptic input. The plots show the stationary firing rate as a function of $f_{in}$, the firing rate of excitatory presynaptic inputs. {\em Dots:} firing rate from simulations of Eq.~\ref{eq:V} with $J_{ij}=1$ and input firing rate reported on the horizontal axis. {\em Line:}  response function under the diffusion approximation, Eq.~\ref{eq:lifrf}.}
\label{fig:lifrf}      
\end{figure}

We note that condition i) is rather realistic in cortex, where values of $J$ are estimated to be about $1/20$ or less of the difference $\theta - V_L$ (for example, $J \approx 0.5$~mV with spike thresholds $10-20$~mV above rest; see e.g. \cite{sn94}). Condition ii), also known as the {\em white noise} approximation, should hold self-consistently in the whole network and for finite values of the synaptic weights it remains an approximation \cite{l06,Pena:2018ui,vellmer2019} (we'll say a bit more on this in Sec.~\ref{sec:corr}). For feedforward input, however, the diffusion approximation gives excellent results for the firing rate of the LIF neuron, as shown in Fig.~\ref{fig:lifrf}. The response function shown in the figure reads \cite{j68,at92,ab97}
\be \label{eq:lifrf}
\Phi(\mu,\sigma)= \left ( \tau_r + \tau \sqrt{\pi} \int_{V_r-\mu \over \sqrt{2} \sigma}^{\theta-\mu \over \sqrt{2} \sigma} dx e^{x^2} (1+\erf(x)) \right )^{-1},
\ee
where $\erf(x)$ is the error function and $\mu, \sigma$ are given by Eqs.~\ref{eq:mu-sigma-2}. As a reminder, the error function is defined as
\be \label{eq:erf}
\erf(x) = { 2 \over \sqrt{\pi} } \int_0^x dz e^{-z^2} =  2  \int_0^{\sqrt{2}x} { dz \over \sqrt{2 \pi} } e^{-{z^2 \over 2}} = P(|z| < \sqrt{2} x),
\ee
where $z \sim \mathcal N(0,1)$ is a standard Gaussian random variable. Two different derivations of Eq.~\ref{eq:lifrf} can be found e.g. in \cite{j68} and \cite{b00} (see also \citenoparens{s51}). Fig.~\ref{fig:lifrf} shows that Eq.~\ref{eq:lifrf} is in excellent agreement with simulations despite a finite $J$ (in figure, $J=1$ and $\theta-V_L=20$). It has also been determined experimentally that this response function describes quite accurately the response function of real cortical neurons \cite{lgsf08}. 

\subsection{The mean field equations}

With the response function in hand, we can write the self-consistent mean field equations for populations of LIF neurons (in vectorial notation):
\be \label{eq:mfLIF}
{\pmb f} = {\bf \Phi} ({\pmb \mu({\pmb f}), {\pmb \sigma}({\pmb f})}),
\ee
where $\Phi_i$ is given by \ref{eq:lifrf} and $\mu_i, \sigma_i$ are given by \ref{eq:mu-sigma-2}. In the more general case of $M$ clusters with random connectivity, Eqs.~\ref{eq:mu-sigma-2} generalize to (see Sec.~\ref{sec:randconn})
\begin{equation} \label{eq:c-mu-sigma-4}
\mu_{\alpha} = V_{\alpha,L}^*+ \sum_{\beta=1}^M c_{\alpha \beta} N_{\beta} [J]_{\alpha \beta} f_{\beta} \tau_{\alpha}, \quad \sigma^2_{\alpha} = { 1 \over 2 } \sum_{\beta=1}^M c_{\alpha \beta} N_{\beta} [J^2]_{\alpha \beta} f_{\beta} \tau_{\alpha},
\end{equation}
where $ [J]_{\alpha \beta}<0$ if $\beta$ is an inhibitory population, and some terms may reflect a {\em synaptic} input coming from external pools of neurons (see Eqs.~\ref{eq:muext}). For simplicity, we have assumed the same equilibrium value for all neurons in the same population ($V_{\alpha,L}^* \doteq V_L+\tau I_{\alpha,ext}$), a restriction that can be easily removed. Note the similarity of Eqs.~\ref{eq:c-mu-sigma-4} with the relations~\ref{eq:c-mu-sigma} valid for the binary neuron, and note that \ref{eq:c-mu-sigma-4} hold for the membrane potential. For the input current, Eq.~\ref{eq:Ilif}, expressions identical to \ref{eq:c-mu-sigma} hold (see appendix~\ref{app:mu-sigma} for details). 

Although we have outlined the theory for networks of LIF neurons, the same theory applies to networks of other integrate-and-fire neurons, such as the quadratic and the exponential integrate-and-fire neuron, and even to some conductance-based models. The main difference is in the response function to be used; see e.g. \cite{fm99,fhvb03,r04,adapt04}.

%

\vsp Spiking networks with inhibitory and excitatory populations can exhibit a repertoire of different behaviors. They can produce fast global oscillations with asynchronous spiking in single neurons, globally synchronized states as well as states of asynchronous activity (e.g., \cite{vs98,bh99,b00,g00,md02}). In the presence of multiple clusters, complex network configurations are possible \cite{Mazzucato:2015jk}, including configurations wherein an excitatory cluster is active on the backdrop of a globally spontaneous activity state. Such a model was first introduced and analyzed by~\citetext{ab97} with the mean field approach outlined here, and was proposed as a biologically plausible model of working memory capable of storing and retrieving multiple memories. In particular, if single neurons can code for multiple stimuli (as observed in real cortical neurons), an extensive number of stimuli can be accommodated \cite{cmla04}. These networks are rather complex but still amenable to a mean field analysis which is a direct generalization of the approach outlined here. Networks of this kind have also been proposed as mechanistic models of decision making in cortical circuits, see e.g. \cite{wang:2008}. More recently, similar models have been used explain the emergence of slow fluctuations in firing rates. We discuss them in Sec.~\ref{sec:finitesize}. 

\section{Validity of the mean field approximation} \label{sec:validity}

In the mean field procedure carried out in Sec.~\ref{sec:mfmain}, we have assumed that the sum over many independent inputs causes the input current $I_i$ to be distributed as a Gaussian distributed variable. We have then neglected the fluctuations of $I_i$. By doing so, we have assumed that each neuron receives the {\em mean input} generated by the other neurons and by the external currents. In Sec.~\ref{sec:extensions} we have included the effect of the Gaussian fluctuations into the theory, as well as the effect of random connectivity. In this picture, each neuron receives a Gaussian input current with given mean and variance that depend, self-consistently, on the activities of the other neurons.

In this section we consider the assumptions made so far and some of the consequences of violating those assumptions. 

\subsection{Implications of the thermodynamic limit. Synaptic scaling} \label{sec:scaling}

We first note that the theory applies only to large networks. Only when the input is the sum over many independent terms we can apply the central limit theorem and replace the input current with a Gaussian variable. Therefore, this assumption requires to perform the thermodynamic limit $N \to \infty$, which in turn implies that the synaptic weights must be scaled with $N$ for the input current to remain finite in the limit.  

For concreteness, we consider a single population of binary neurons with constant external current. The input is a sum of $N$ terms of order 1; assuming independent contributions from presynaptic neurons, both the mean and the variance will grow as $N$ (neglecting $I_{i,ext}$ for now, we focus on the recurrent contributions): 
\be
\langle I_i \rangle  = \langle \langle \sum_j^N J_{ij} x_j \rangle \rangle \sim \mathcal O(N), \quad \text{Var}(I_i) = \text{Var}(\sum_j^N J_{ij} x_j) \sim \mathcal O(N),
\ee
where the symbol $\mathcal O(N)$ means ``order $N$" as $N \to \infty$, i.e., $\mathcal O(N)/N \to constant$. Receiving a large input, all neurons will saturate to their maximal activity value, which would render the state of the network useless for computation. Therefore, we must rescale the synaptic weights so as to produce a finite activity.

\vsp Two main scaling options have been used: one is to scale the weights as $J_{ij} \sim g/N$ and the other is to scale them as $J_{ij} \sim g/\sqrt{N}$. 
\begin{itemize}

\item With the first choice, $J_{ij} \sim g/N$, we obtain 
\be
\langle I_i \rangle  \sim \mathcal{O}(1), \quad \text{Var}(I_i) \sim {1 \over N}.
\ee
The variance vanishes in the limit. In this case, the input current ceases to fluctuate, and all neurons receive exactly the same input and therefore will have the same neural activity. The mean field prediction in this case is $f_i = f$ for all $i$, with $f$ given by the self-consistent Eq.~\ref{eq:meanfieldhom},
\be
f = {\mathcal S}(g f + I_{ext}), 
\ee
analyzed in Sec.~\ref{sec:hom}.

\item With the second choice, $J_{ij} \sim g/\sqrt{N}$, we have 
\be
\langle I_i \rangle   \sim \mathcal{O}(\sqrt{N}), \quad \text{Var}(I_i) \sim \mathcal O(1).
\ee
In this case, the mean input tends to increase with $N$ while the variance remains finite (``order $1$"). The prediction is that all neurons' activities will saturate to their maximal value, as in the absence of scaling. This can be avoided by adding inhibitory populations of neurons, as shown next.

\end{itemize}

\subsubsection{Balanced networks}

The problem encountered with the $1/\sqrt{N}$ scaling may be rescued by inhibition. Let's consider a network with one excitatory and one inhibitory population, each of size $N$. For simplicity, we consider constant synapses in each population. By rescaling all weights as $J_{{\alpha}{\beta}} = \tilde J_{{\alpha}{\beta}}/\sqrt{N}$ and the external current as $\mu_{E,ext} \sim \tilde \mu_{E,ext} \sqrt{N}$, Eqs.~\ref{eq:mu-sigma} give:
\begin{eqnarray} \label{eq:balance0}
\mu_E & = & \sqrt{N} (\tilde J_{EE} f_E - \tilde J_{EI} f_I + \tilde \mu_{E,ext}) \\ \label{eq:balance0b}
\mu_I & = & \sqrt{N} (\tilde J_{IE} f_E - \tilde J_{II} f_I + \tilde  \mu_{I,ext}).
\end{eqnarray}
In the same limit, the variances remain finite: 
\begin{eqnarray} \label{eq:var}
\sigma^2_E & = & \tilde J_{EE}^2 f_E + \tilde J_{EI}^2 f_I \\ \label{eq:varb}
\sigma^2_I & = & \tilde J_{IE}^2 f_E + \tilde J_{II}^2 f_I. 
\end{eqnarray}
Now we require that the mean inputs remain $\mathcal O (1)$ in the limit of large $N$. This requires the inhibitory and excitatory components in Eqs.~\ref{eq:balance0}-\ref{eq:balance0b}  to cancel out in the limit, at least within order $1/\sqrt{N}$:
\begin{eqnarray} \label{eq:balance}
\tilde J_{EE} f_E - \tilde J_{EI} f_I + \tilde \mu_{E,ext} & = & { \mu_E \over \sqrt{N}} \to 0 \\ 
\tilde J_{IE} f_E - \tilde J_{II} f_I + \tilde \mu_{I,ext} & = & { \mu_I \over \sqrt{N}} \to 0,
\end{eqnarray}
or, in matrix notation,
\be \label{eq:balance1}
\tilde J f = - \tilde \mu_{ext}.
\ee
Networks with this property are known as `balanced' networks \cite{vs96,vs98} and are the object of much research because they share many properties of real cortical circuits, including erratic spike trains that are difficult to explain without the balance hypothesis (see the next section). Note that the input current retains its variability in the thermodynamic limit, justifying the introduction of $\sigma$ in the extended mean field theory of Sec.~\ref{sec:variance} (an alternative justification would be the presence of an external fluctuating input for any $N$ \cite{ab97a,ab97}, in which case the synapses can be scaled as $1/N$). 

\begin{remark} -- \label{r:K} {\em We must observe that cortical neurons are not connected to all other neurons in their neural circuit, but they are connected to, say, $K \ll N$ neurons, where $K$ can be large. Therefore, the theory outlined above can be made more realistic by assuming random connectivity with mean $c=K/N$, in which case Eqs.~\ref{eq:c-mu-sigma} hold. In the limit $N\to \infty, K \to \infty$ with $K/N \to c$ and $J_{\alpha \beta} \sim 1/\sqrt{K}$, from the above equations we see that the mean will grow as $\sqrt{K}$ while the variance will remain finite. All the arguments remain the same, except that we replace $N$ with $K$. In this version of the theory, synapses are required to scale as the inverse of the square root of their mean number of afferents $K$, rather than the total number of neurons. Partial evidence for a $1/\sqrt{K}$ scaling of cortical synapses has been reported in \cite{Barral:2016nt}.}
\end{remark}

\begin{remark} --  \label{rem:balance} {\em We should notice that the argument of the previous remark (and Eqs.~\ref{eq:balance0}-\ref{eq:varb} on which it is based) holds only if one can prove that correlations vanish in the thermodynamic limit. Technically, this requires $K < \ln N$ \cite{dgz87}, which in turn implies $c=0$ in the thermodynamic limit. However, it turns out that, in a balanced network, this condition is not necessary \cite{Renart:2010pb}. See also Sec.~\ref{sec:corr}.}
\end{remark} 

\subsubsection{Mean field theory of balanced networks}

The mean field theory of the balanced network proceeds as follows: the balanced solution $f^*$ is given by the solution to Eq.~\ref{eq:balance1}, i.e.
\be \label{eq:balance2}
f^* = - \tilde J^{-1} \tilde \mu_{ext}.
\ee
This is a necessary condition for the existence of the balanced state; additional conditions must be imposed to guarantee positive, non-saturating firing rates \cite{vs98}. In a recurrent network, we have the additional requirement that the output rate of a neuron in a population must match its own input firing rate:
\begin{eqnarray}
f_E & = & \Phi_E(\mu_E(f_E,f_I), \sigma_E(f_E,f_I)) \\
f_I & = & \Phi_I(\mu_I(f_E,f_I), \sigma_I(f_E,f_I)),
\end{eqnarray}
where $\Phi_{\alpha}$ is the response function of the neurons in the $\alpha$ population. Note note that $f_E^*$ and $f_I^*$ given by Eqs.~\ref{eq:balance2} provide a value for the variances (according to Eqs.~\ref{eq:var}-\ref{eq:varb}), but do not provide a value for the input means $\mu_{E,I}$ defined in Eqs.~\ref{eq:balance0}-\ref{eq:balance0b} (these are of order 1 but are not necessarily zero, even in the thermodynamic limit). Therefore, one imposes the balance condition and derives $\mu_{E,I}$ self-consistently: 
\begin{eqnarray}
\mbox{find $\mu_E$, $\mu_I$ so that:} \quad f_E^* & = & \Phi_E(\mu_E, \sigma_E(f_E^*,f_I^*)) \\
f_I^* & = & \Phi_I(\mu_I, \sigma_I(f_E^*,f_I^*)).
\end{eqnarray}
Note that $\mu_{E,I}$ depend on $f_{E,I}^*$: for example, if the external currents are varied, one obtains new $f_{E,I}^*$ values and thus new $\mu_{E,I}$ from the self-consistent equations above.

\vsp We make a few more important remarks regarding balanced networks:

\begin{itemize}

\item Eq.~\ref{eq:balance1} shows that the balanced state requires an external current, and the external current must be of order $\sqrt{K}$ (see Remark~\ref{r:K}). Without an external current, Eq.~\ref{eq:balance1} reads $\tilde Jf=0$. This case is problematic in several ways. For example, when a non-zero solution exists for the firing rates, the latter could have large fluctuations in the null subspace of $\tilde J$, and the asynchronous state could be lost. Mean field theory with a singular synaptic matrix is, in general, problematic.

\item Eqs.~\ref{eq:balance2} imply that the balanced rates depend linearly on the external input current, which is at odds with the bistability studied in Sec.~\ref{sec:bist} resulting in macroscopic changes in firing rates. In other words, a network cannot be balanced and bistable at the same time \cite{rmwp07}. To obtain a bistable balanced network, other sources of non-linearity must be leveraged, such as short term plasticity \cite{Barbieri:2007zl,Mongillo:2012px}. 

\item In a balanced network, rates dynamically adjust to balance excitatory and inhibitory inputs, so that the mean and the variance of the input remains of order 1 \cite{vs98}. When the mean input is below threshold, firing is due to input fluctuations and this regime is stable for continuous perturbations of the input. This produces erratic spike trains without the need for fine tuning. 
\end{itemize}

\subsection{The role of correlations} \label{sec:corr}

Correlations of neural activity come in two main flavors, spatial (called {\em cross-correlations}) and temporal (called {\em autocorrelations}). Mean field theory makes specific assumptions about them: in the most basic form, both forms of correlations are supposed to vanish in the thermodynamic limit. We briefly discuss the role of correlations in this section.

\subsubsection{Spatial correlations}

The application of the central limit theorem invoked in Sec.~\ref{sec:scaling} also requires negligible correlations between the activities of the neurons. For example, if the synaptic weights are symmetric, $J_{ij}=J_{ji}$, then the random variables $J_{ij} x_j$ and $J_{ji} x_i$ could be correlated. If the variables are highly correlated, global oscillations of the firing rates may emerge, and the asynchronous regime is lost. Sparse connections typically reduce the correlations between neurons, and are very often invoked (see below); other mechanisms, such as the balance of excitation and inhibition discussed in the previous section, are effective at reducing correlations even in networks that are not sparse \cite{Renart:2010pb,Helias:2014hl}.

\begin{defn}[sparseness] A network is sparse when its neurons receive a mean number of connections $K \ll N$, such that the average connectivity $c={K \over N} \to 0$ as $N \to \infty$. 
\end{defn}

Note that the definition above does not exclude the possibility that $K \to \infty$ in the thermodynamic limit. In fact, in the theory of balanced networks, we take both $N \to \infty$ and $K\to \infty$. In finite networks, where $K$, $N$ and $J$ are all finite, sparseness becomes a messier concept. A more useful approach in that case might be to specify the conditions on the values of $K$ and $J$ resulting in negligible correlations between the spike trains coming from different neurons. Normally, in this regime mean field theory will be quite accurate.

In the presence of correlations, some of the formulae derived earlier may not hold. For example, consider Eq.~\ref{eq:meanI2} and its generalization to random connectivity, Eq.~\ref{eq:meanI2c}: by using the general fact that $\langle y x \rangle = \langle y \rangle \langle x \rangle + \text{Cov}(y,x)$, we get (with $\langle y \rangle = \langle c_{ij} J_{ij}\rangle = c [J]$):
\begin{eqnarray}
\langle \langle \sum_j^N c_{ij} J_{ij} x_j \rangle \rangle & = & c N [J] f + c  \sum_j^N \text{Cov}(J_{ij} x_j) \\
& = & K [J] f + K \langle \text{Cov} \rangle,
\end{eqnarray}
where $c=K/N$ and $\langle \text{Cov} \rangle \doteq N^{-1} \sum_j^N \text{Cov}(J_{ij} x_j)$. We see that if the mean covariance does not vanish in the limit, the two terms on the right hand side have the same order of magnitude. A similar argument applies to the sum over the presynaptic neurons of quantity \ref{eq:varJx}. In this case we have to use the more general formula for the variance of $\sum_j^N c_{ij} J_{ij} x_j$:
\be \label{eq:cov}
\text{Var}(\sum_j^N c_{ij} J_{ij} x_j) = \sum_j \text{Var}(c_{ij} J_{ij} x_j)+ 2 \sum_{j<k} \text{Cov}(c_{ij}J_{ij} x_j ,c_{ik} J_{ik} x_k).
\ee
Note that the first term of the right-hand side of Eq.~\ref{eq:cov} is a sum over $N$ terms, whereas the second is a sum over $\mathcal O(N^2)$ terms, and therefore the second term may not be negligible compared to the first. 

It must be noted that introducing scaling laws for $c=K/N$ and $J_{ij}$ in these formulae may not be sufficient to determine the impact of the covariance terms. For example, if the network is in the asynchronous regime, the covariance terms vanish by definition. In general, several ingredients in addition to connectivity contribute to the degree of cross-correlations between spike trains in a recurrent network of spiking neurons (see e.g. \cite{Ostojic2009-ee}), and each case may have to be analyzed separately. Two important examples in which cross-correlations vanish in large networks are balanced networks (even dense ones \cite{Renart:2010pb,Helias:2014hl}) and networks with $K < \ln N$ \cite{dgz87}.

\subsubsection{Temporal correlations}

Another important assumption of the theory is the absence of temporal correlations in the activity of single neurons, as quantified by their autocovariance (AC). Whereas the firing rate is the average of the activity, e.g. $\langle x_i \rangle$, the AC at lag $\tau$ is given by 
\be
AC_i(\tau) = \langle x_i(t) x_i(t+\tau) \rangle -  \langle x_i(t) \rangle \langle x_i(t+\tau) \rangle.
\ee
Often one computes the autocorrelation instead, which is just a normalized version of the AC. The AC is a measure of the similarity between the activity of a neuron at two time points. For example, the AC of $x(t)=\cos(\omega t)$ is itself a cosine function. In the asynchronous regime with stationary firing rate, the AC is a delta function,
\be
AC_i(\tau) = AC_i(0) \delta (\tau).
\ee
We refer to this assumption as the `white noise' approximation. The presence of temporal correlations poses two main problems to the theory: 

\begin{itemize}
\item {\bf Determination of the firing rates.} When the activity depends only on the current value of the input, as in our network of binary logistic neurons, the autocorrelation does not affect the determination of the firing rates. The activity of LIF neurons, however, depends on previous history at least back to the time of their previous spike. The response function of the LIF neuron (Eq.~\ref{eq:lifrf}) correctly describes the firing rate only for white noise input. It is often argued that in a large network, the sum of many input spike trains will converge to a delta-correlated input current, but in general this is not strictly correct \cite{Lindner:2006zl,Cateau:2006tu,mrp08}. The temporal correlations present in the input spike trains may therefore survive in a large network. Examples of such correlations in spiking neurons are due to a finite refractory period, which introduces a negative AC at very short lags, the finite rise and decay time of receptor-mediated current, and firing rate adaptation. Improved response functions in the presence of synaptic filtering have been found, e.g.  \cite{fb02,mp04}. However, the recurrent nature of the network may induce finite correlation times in a network that otherwise has not built-in temporal correlations \cite{Fulvi-Mari:2000fp,l06}, which leads us to the next point.

\item  \vsp {\bf Self-consistent theory of correlations.} Aside from the firing rates, a satisfactory theory should also determine self-consistently the autocovariance of the activity of a recurrent network. Self-consistent descriptions of AC have been obtained with a variety of methods, some also applicable to spiking networks \cite{Sompolinsky1988-tn,l06,Harish2015-kp,Mastrogiuseppe2017-qi,Pena:2018ui,vellmer2019}. These efforts have shed light on the dynamical behaviors of neural networks, as well as the transitions among them, as one or a few key parameters are varied.
\end{itemize}

\subsection{Finite size effects} \label{sec:finitesize}

The theory requires the thermodynamic limit $N \to \infty$. However, it typically works well also in finite networks, as confirmed by the agreement with numerical simulations. In a finite network, however, discrepancies from the mean field predictions can be observed. Fluctuations in the network's activity can destabilize fixed points that would otherwise be stable in the infinite network (for some relevant applications, see e.g. \cite{Miller:2006mb,Braun:2010uq}). In this section, we briefly discuss two possible consequences of having a finite number of neurons: a spatial variation of firing rates, and metastability. 

\subsubsection{Spatial variation of firing rates}

Mean field theory assumes that all neurons of a homogeneous population have the same firing rate. Due to random connectivity, neurons will receive input from a mean number of $K=cN$ neurons, with variance $c(1-c)N = K(1-K/N)$. The cell-to-cell fluctuations in the number of inputs scale therefore as $\sqrt{K}$. When $K$ is large, the fluctuations are negligible compared to the mean, so that all neurons receive the same fraction of inputs, $c$ (the distribution of $k/N$, where $k$ is the number of connected inputs, converges to a $\delta$ function centered in $\lim_{N \to \infty} K/N=c$). In a finite network, however, fluctuations in the number of inputs can induce variability in the firing rates across neurons. This is especially true in a balanced network, where the mean input scales as $\sqrt{K}$, the same order of magnitude of the spatial fluctuations \cite{vs98}. Both in these models and real cortical circuits, the spatial distributions can be quite wide, but are well predicted by a mean field analysis that treats the distribution of firing rates self-consistently, see e.g. \cite{ab97a}.  

\subsubsection{Metastability} 

\begin{figure}
\centering
\includegraphics[scale=.48]{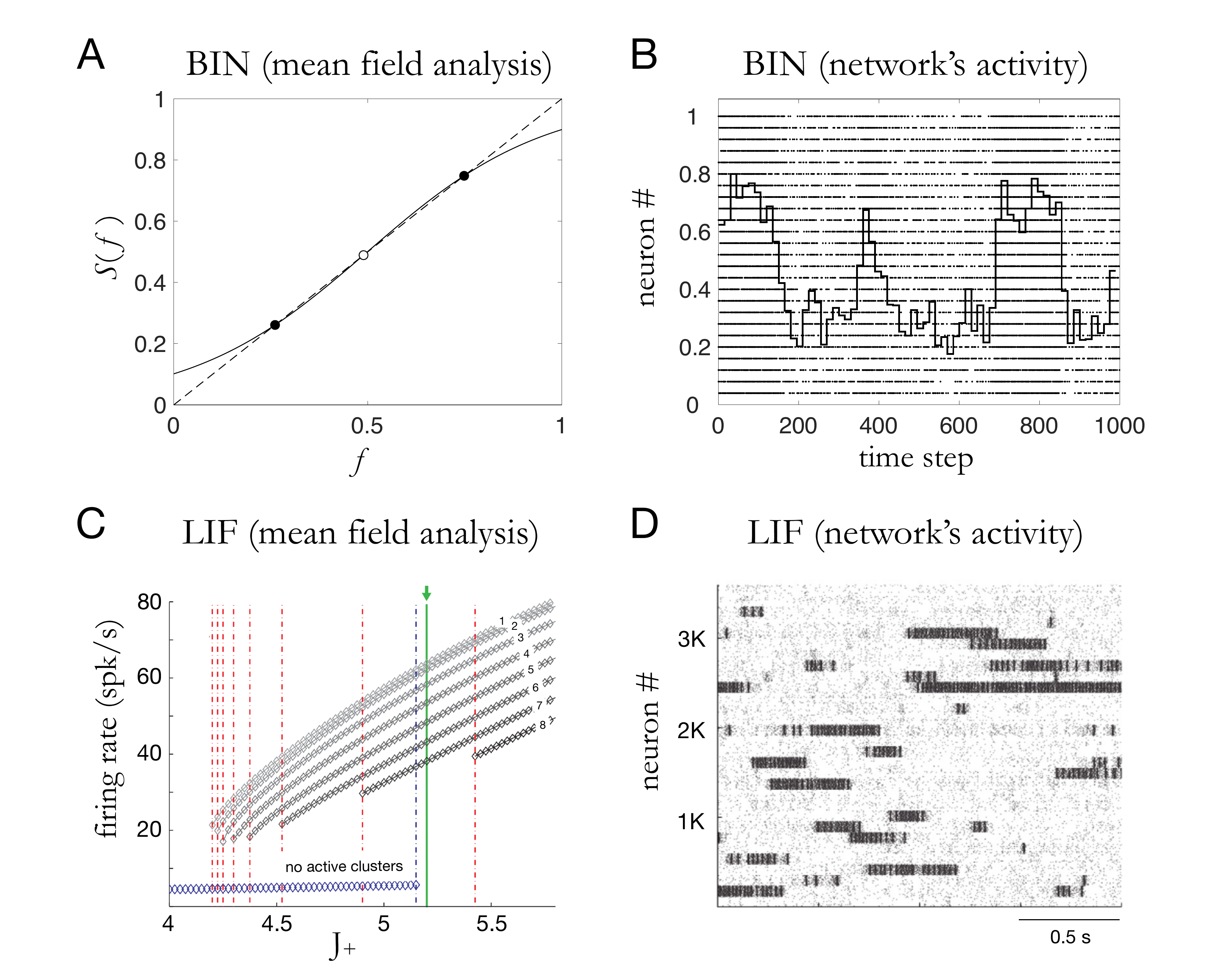} 
\caption{\small Metastability due to finite size effects. {\bf A.} Graphical solution of the mean field equations for the binary network of Fig.~\ref{fig:bist} with $g=1.1$, $I_{ext}=-0.6/g$. {\bf B.} Rasters and ensemble average of the network of panel A with $N=100$ neurons. Finite size effects cause the activity of the network to flip among the two stable fixed points shown in A (black circles; these fixed points are stable in the infinite network). {\bf C.} Mean field analysis of a clustered network of LIF neurons with $30$ excitatory clusters, analogous to the bifurcation diagram of Fig.~\ref{fig:bist}B. In this case there are multiple upper branches, each characterized by a different number of simultaneously active clusters (from $1$ to $8$). A new brach appears as soon as the relative potentiation of synaptic weights inside clusters ($J_+$) crosses a critical point (vertical red lines). {\bf D.} Raster plot of the network of panel C for $J_+=5.2$ (green vertical line) showing rich metastable dynamics. Note that this network is completely deterministic. Panels C and D adapted from~\protect \citetext{Mazzucato:2015jk}.}
\label{fig:metastable}      
\end{figure}

The spatially distributed firing rates mentioned in the previous paragraph tend to be stable despite the finite size of the network. A different phenomenon is metastability, where the firing rates in subpopulations of neurons are homogeneous and well predicted by mean field theory, but the activity of the network is not stationary. Metastability occurs when the stable fixed points of activity are destabilized by fluctuations due to finite $N$. For metastability to occur one needs at least two stable fixed points which lose stability in the finite network. This case is illustrated in the Fig.~\ref{fig:metastable}A-B for the network of Fig.~\ref{fig:metastable} with $g=1.1$, $I_{ext}=-0.6/g$ and $N=100$. When this network has fewer than $1,000$ neurons, there are enough fluctuations to cause the network's activity to randomly flip between the two fixed points shown in panel A. The larger $N$, the longer the time spent in each point.

%


In a network partitioned in many clusters and including recurrent inhibition, a mean field analysis shows the existence of a large variety of fixed points \cite{Mazzucato:2015jk}, and the interplay of recurrent inhibition and finite size fluctuations brings about a rich metastable dynamics \cite{Deco2012-wg,Litwin-Kumar:2012ty,Mazzucato:2015jk}, as shown in Fig.~\ref{fig:metastable}C-D.

This occurs when the mean synaptic weights inside the excitatory clusters are strong enough, with the critical point being accurately predicted by mean field theory. In fact, there are a multitude of critical points for the mean synaptic weights, as shown by the vertical red lines in Fig.~\ref{fig:metastable}C. Above the smallest critical point, only one excitatory cluster can be active at any given time. Above a second critical point, up to $2$ clusters can be active, and in general beyond the $n^{th}$ critical point, up to $n$ clusters can be active. See \cite{Mazzucato:2015jk} for details. 

\section{Discussion and conclusions}

In this chapter, we have presented an elementary introduction to the mean field approach for populations of spiking neurons. This is an approach borrowed from physics that allows to study the behavior of large networks by replacing the input to a neuron with a mean field generated by its afferent neurons. A key feature is self-consistency, which has two meanings: in the first, it means that the properties of input and output neurons in homogeneous populations must match; in the second, it means that the conditions assumed {\em ab initio} to develop the theory (such as Gaussian current) must indeed occur. In the most basic setting, we focus on matching self-consistently the firing rates and neglect other important properties such as the autocorrelations of the neurons. The theory, however, can be generalized to include those properties as well. 

It may seem strange that a theory based on neglecting fluctuations or assuming stationary behavior can give useful predictions in the presence of both fluctuations and temporal dynamics. Yet we are familiar with the success of such theories in physics where, in studying large systems composed of interacting particles, thermal fluctuations will cause the particles to move around or to flip their spins while macroscopic properties of the system, such as volume, pressure or energy, may remain constant. 

The theory is, strictly speaking, only valid in the thermodynamic limit and under restrictive conditions, such as stationary activity and low correlations among neurons. However, it can be extended in a number of ways, for example by including the fluctuations of the input current or determining, self-consistently, the spatial variation of firing rates across the neurons of a finite network. The theory has allowed an understanding of the behavior of networks partitioned in populations of excitatory and inhibitory neurons, each of which can be further partitioned in sub-clusters of different cell types along the lines discussed in Sec.~\ref{sec:randconn} and Sec.~\ref{sec:finitesize}.

There are many other ways in which the theory can be extended. By looking at the self-consistent autocorrelation in mean field, it was found that rate models such as Eqs.~\ref{eq:meanIeq} can exhibit deterministic chaos in the dynamics of the firing rates \cite{Sompolinsky1988-tn,Molgedey:1992tt,Rajan2010-pf,Kadmon2015-cb,Aljadeff:2015zm}. In this context, the mean field approach is more commonly referred to as `dynamical mean field theory' \cite{cs18,Schuecker:2016gf}. Similar efforts are being carried out in spiking networks, where the possibility of firing rate chaos is still an open question \cite{Ostojic2014-qe,Wieland:2015wq,Harish2015-kp}. Other ways in which the theory can be extended include ways to determine self-consistently the effects of firing rate adaptation \cite{t93,adapt04,gmd07}, neuromodulators \cite{bw01}, temporal and spatial correlations \cite{Ginzburg:1994eq,Meyer:2002yl,l06,vellmer2019}, short-term plasticity \cite{Barbieri:2007zl,Mongillo:2012px}, voltage-dependent conductances \cite{Kumar:2008vn,Capone:2019ne,Sanzeni:2020bx}, spatial topology \cite{wc73,Pyle:2017gq}, and more. Within these extensions, mean field theory has been applied to increasingly more realistic models of neural activity.

Mean field theory can also be useful when its assumptions are violated, as e.g. in metastable networks. These networks can have a large number of stable configurations for $N \to \infty$, which give rise to rich metastable dynamics in the case of a network of finite size. This type of metastable dynamics has been found in the neural activity of humans and other behaving animals \cite{Miller2016-kx,La-Camera:2019zm}. Mean field theory allows to locate the metastable regime on a bifurcation diagram, and accurately predict the firing rates of neural clusters \cite{Mazzucato:2015jk} during metastable activity. Metastable networks can also explain the emergence of slow fluctuations and the quenching of trial-to-trial variability in response to sensory stimulation \cite{Deco2012-wg,Litwin-Kumar:2012ty}. 

Some of the studies mentioned above were actually performed using a version of mean field theory known as the population density approach \cite{k72,t93,av93,fm99,bh99,nt00,n00,b00,md02,vellmer2019}. In this approach, one obtains not only the stationary distribution of the firing rates, but also the distribution of the membrane potentials. By leveraging perturbative solutions of a Fokker-Planck equation, this approach can help uncover the dynamics emerging from the instability of the asynchronous regime, and it has been used to build complete phase diagrams of networks of spiking neurons \cite{bh99,b00}. In some special cases, exact results in the thermodynamic limit for the dynamics of both the firing rate and the membrane potential have been obtained \cite{roxinPRX15}. Mean field theory, dynamical field theory and the population density approach are complementary approaches, and one may choose one approach or the other according to the problem at hand. 

In conclusion, models of neural circuits are complex dynamical systems capable of a large repertoire of behaviors. Mean field theory, in his diverse incarnations, is one of the few tools at our disposal (and arguably the most successful) at predicting the collective behavior of such models. As we learn more about the potential link between neural dynamics and brain function, these methods are being rediscovered and sharpened to deal with increasingly more sophisticated applications. We hope that this elementary introduction can be useful as a first exposure to the ideas and methods of this approach as applied to networks of spiking neurons, while referring e.g. to \cite{a89,Hertz:1991lr} for comprehensive treatments in the field of neural computation, and to \cite{dfm03,rbw03ch,hertzChapter04,gerstner2book2014} for applications to spiking neurons and related topics.

\section{Acknowledgements}

The author is indebted to Dr. Gianluigi Mongillo for many helpful discussions during the gestation of this chapter and for useful comments on an earlier version of the manuscript. Many thanks also to Xiaoyu Yang and Dr. Maurizio Mattia for a careful reading of an earlier version of the manuscript and for many useful comments. G.L.C. is supported by a U01 grant from the NIH/NINDS Brain Initiative (1UF1NS115779), a grant from the Human Frontier Science Program (HFSP - RGP0002/2019), and a grant from the Office of the Vice President for Research of Stony Brook University (award 1153707-2-63845). The content of this article is solely the responsibility of the author and does not necessarily represent the official views of the National Institutes of Health, the Human Frontier Science Program, or Stony Brook University.

%
\section{Appendix}

\subsection{Ensemble average}\label{app:EA}

To measure the stationary firing rates observed in simulations we can use  Eq.~\ref{eq:ftimeaver}, but when all neurons have the same firing rate we gain precision by using an ensemble average (EA) across the whole population. Due to the average across neurons, the EA can be accurate also in small time bins, providing a time-dependent measure of firing rate as shown in e.g. Fig.~\ref{fig:hom}A (thick line superimposed to the rasters). 

More formally, the EA associates each small time bin $\Delta_t = (t,t+\Delta t)$ with the mean spike count across the population of neurons in that bin:
\be
EA(t)={1 \over N \Delta t} \sum_{k=1}^N n_k(t),
\ee
where $N$ is the number of neurons in the population and $n_k(t)$ is the spike count of neuron $\# k$ in bin $\Delta_t$. 

As $N \to \infty$ and $\Delta t \to 0$ (or $\Delta t \to 1$ in the case of discrete time dynamics), the EA converges to the mean instantaneous firing rate of each neuron inside the population, tracking accurately changes in firing rate over time. In the case of stationary asynchronous activity, the EA for $N \to \infty$ is by definition a flat function of time. This is evident from Fig.~\ref{fig:hom}A although, due to finite $N$, bin by bin fluctuations around the mean are also visible. The stationary firing rate of any neuron over a time interval $T \doteq n \Delta t$ can be estimated through the temporal average of the EA: ${1 \over n} \sum_{t=0}^{n-1} EA(t)$. 

\vsp Note that the EA defined here is closely related to the so-called peristimulus time histogram (PSTH), a widely used measure of neural activity. The difference between the two is that in the PSTH, instead of the spike trains of $N$ neurons recorded in the same trial, one has the spike trains of $1$ neuron recorded over $N$ trials of the same kind (e.g., in response to the same stimulus), with all trials being aligned to the same reference event time (e.g., stimulus onset). 

\subsection{Mean and variance for the LIF neuron} \label{app:mu-sigma}

To main goal of this appendix is to derive Eqs.~\ref{eq:mu-sigma-2}. We shall derive the more general Eqs.~\ref{eq:c-mu-sigma-4}. To goal is to compute the mean and variance of (see Eq.~\ref{eq:Vsol})
\be \label{eq:Vsol1}
V_i(t) = V_i(0) e^{-t /\tau} + \sum_j^N \sum_k^{N^j_t} J_{ij} e^{-(t-t^j_k)/\tau} \Theta(t-t^j_k),
\ee
where $N^j_t$ is the number of spikes arriving in the interval $(0,t)$ from presynaptic neuron $j$. We assume Poisson, independent presynaptic spike trains with firing rates $f_j$. Note that to simplify the upcoming formulae we have identified $V_i$ with $V_i-V_L-\tau I_{i,ext}$, or equivalently, we have set $V_{i,L}^*=V_L+\tau I_{i,ext}=0$. The value of $V_{i,L}^*$ does not affect the variance but must be added to the mean $\mathbb E(V)$ derived below.

\subsubsection{Mean of $V_i(t)$}
Recall that $J_{ij}$ is a random variable. Since the $N^j_t$ are themselves random variables and $t$ can take any positive value, we must use Wald's identities (see e.g. the appendix of \cite{Soula:2006jk}) to determine the mean and variance of $V_i(t)$. The first Wald identity states that, if $N_t$ and $X_k$ are random variables with finite means, then $\mathbb E(\sum_k^{N_t} X_k) = \mathbb E(N_t) \mathbb E(X_k)$. We need to apply this identity to the sum 
\be
\sum_j^N \sum_k^{N^j_t} J_{ij} e^{-(t-t^j_k)/\tau} \Theta(t-t^j_k)  = \sum_{t^j_k}^{N_t} J_{ij} e^{-(t-t^j_k)/\tau} \Theta(t-t^j_k), 
\ee
where we have defined the total number of spikes $N_t = N^1_t + N^2_t + ... + N^N_t$. 

To lighten the notation, we shall use the symbol $\Theta_k$ to mean $\Theta(t-t^j_k)$. Applying Wald's identity we get
\be
\mathbb E(\sum_{t^j_k}^{N_t} J_{ij} e^{-(t-t^j_k)/\tau} \Theta_k) = \mathbb E(N_t) \; \mathbb E(J_{ij} e^{-(t-t^j_k)/\tau} \Theta_k).
\ee
During the interval $(0,t)$ the neuron receives $N$ Poisson spike trains with rate $f$, hence
\be
\mathbb E(N_t) = N f t.
\ee
Moreover, 
\be
\mathbb E(J_{ij} e^{-(t-t^j_k)/\tau} \Theta_k) = \mathbb E(J_{ij}) {1 \over t}  \int_0^t du \; e^{-u/\tau} = [J] {1 \over t} \tau (1-e^{-t/\tau}),
\ee
where $[J]$ is the mean with respect to the quenched distribution of the synaptic weights. We conclude, taking the product with $E(N_t)$,
\be
\mathbb E (V_t) = N [J] f \tau (1-e^{-t/\tau}).
\ee
In the presence of random connectivity (Sec.~\ref{sec:randconn}), $\mathbb E(J_{ij})$ is replaced by $\mathbb E(c_{ij}J_{ij})=\mathbb E(c_{ij}) \mathbb  E(J_{ij})=c[J]$. If $V(0) \neq 0$, we must add the (transient) term $V(0) e^{-t /\tau}$ to the mean. If the initial condition $V(0)$ is a random variable $V_0$ (say, random reset value after a spike), then we add $\mathbb  E(V_0) e^{-t /\tau}$ to the time-dependent mean (note that this term is transient): 
\be \label{eq:EVt}
\mathbb E(V_t) = \mathbb E(V_0) e^{-t /\tau} + c N [J] f \tau (1-e^{-t/\tau}).
\ee
In the stationary case ($t \gg \tau$) we finally obtain, for the mean of the free membrane potential,
\be \label{eq:muapp}
\mu = c N [J] f \tau.
\ee
Adding up the inputs from distinct populations and the constant term $V_{\alpha,L}^*$ (equal for all neurons in population $\alpha$), we obtain the first of Eqs.~\ref{eq:c-mu-sigma-4} (recall that $\tau$ is the membrane time constant of the postsynaptic neuron).

\subsubsection{Variance of $V_i(t)$}
Using similar arguments we can compute the variance:
\begin{eqnarray}
\text{Var}(V_t) & = & \mathbb E(V_t^2) - \mu_t^2 \\
& = & \mathbb E(\sum_{t^j_k,t^{j'}_{k'}}^{N_t N'_t} J_{ij} J_{ij'}  e^{-(t-t_k^j)/\tau} \Theta_k e^{-(t-t_{k'}^{j'})/\tau} \Theta_{k'}) - \mu_t^2 \\
& = & \mathbb E(\sum_{t^{j}_{k}}^{N_t} J_{ij}^2  e^{-2(t-t_k^j)/\tau} \Theta_k) + \mathbb E_{t^{j}_{k} \neq t^{j'}_{k'}} (...) - \mu_t^2.
\end{eqnarray}
If the spike trains $\{ t_k^j\}$ and $\{ t_{k'}^{j'} \}$ are independent, $\mathbb E_{t^{j}_{k}\neq t^{j'}_{k'}} (...) - \mu_t^2=0$, and we are left with
\begin{eqnarray}
\text{Var}(V_t) & = &\mathbb E(\sum_{t^{j}_{k}}^{N_t} J_{ij}^2  e^{-2(t-t_k^j)/\tau} \Theta_k) \\
& = & \mathbb E(N_t) \; \mathbb E(J_{ij}^2) {1  \over t } \int_0^t du \; e^{-2u/\tau} \\
& = & { 1 \over 2} N [J^2] f \tau (1-e^{-2t/\tau}).
\end{eqnarray}
If $V(0)$ is a random variable $V_0$, the total variance is the sum of the variances:
\be \label{eq:VarVt}
\text{Var}(V_t) = \text{Var}(V_0) e^{-2t /\tau} + { 1 \over 2} N [J^2] f \tau (1-e^{-2t/\tau}). 
\ee
Note that during transients, the variance is twice as fast as the mean. In the case of random connectivity, we need to replace $[J^2]$ with $\mathbb E(c_{ij}J_{ij}^2)=c[J^2]$ (since $c_{ij}$ and $J_{ij}$ are independent, and $c_{ij}^2=c_{ij}$). After the transient, the variance converges to
\be \label{eq:sigmaapp}
\sigma^2 = { 1 \over 2} c N [J^2] f \tau,
\ee
and adding up the variances from $M$ homogeneous populations we obtain the second of Eq.~\ref{eq:c-mu-sigma-4}.

\subsubsection{The Gaussian picture}
For large $N$, the free membrane potential follows approximately a Gaussian distribution with the mean and variance computed above. Note that, in all formulae derived above, $t \geq 0$ is the time elapsed since the initial time $0$. Those formulae apply for any other initial time (as long as $t$ means the difference between the current time and the initial time); therefore, the same formulae can be used to derive the moments of any increment $dV_t=V_{t+dt}-V_t$ in a small interval $dt$ conditioned on a (fixed) initial condition $V_0=V_t$ (being fixed, it does not contribute to the variance). This is done by replacing $t$ with $dt$ in Eqs.~\ref{eq:EVt} and \ref{eq:VarVt} and using $1-e^{-n dt/\tau} \approx n dt/\tau$ (with $n$ fixed), obtaining, for the generic neuron $i$,
\be \label{eq:EdV}
\mathbb E(V_{t+dt}-V_t) = (- V_t /\tau + \tilde \mu) dt, \quad \text{Var}(dV_t) = \text{Var}(V_{t+dt} - V_t) = \tilde \sigma^2 dt,
\ee
where 
\be \label{eq:momapp}
\tilde \mu =  c N [J] f, \qquad \tilde \sigma^2 = c N [J^2] f.
\ee
Hence we can write
\be \label{eq:dVt}
dV_t = \left ( - { V_t \over \tau }  + \tilde \mu \right ) dt + \tilde \sigma \sqrt{dt} \; z,
\ee
where $z\sim \mathcal N (0,1)$ is, as usual, a standard Gaussian variable.\footnote{\label{ft:Vt}Recall that in these formulae $V_t$ meant $V_t - V^*_L = V_t - V_L - \tau I_{ext}$; by reintroducing these terms into Eq.~\ref{eq:dVt} we get the more general
\be \label{eq:dVt-gen}
dV_t = \left ( - { V_t - V_L \over \tau }  + \tilde \mu + I_{ext} \right ) dt + \tilde \sigma \sqrt{dt} \; z. \nonumber
\ee
} 
Note that Eq.~\ref{eq:dVt} makes it explicit that the fluctuations of $dV_t$ are proportional to the square-root of time, $\sqrt{dt}$, a well known property of diffusion. Eq.~\ref{eq:dVt} is the precursor of a {\em stochastic differential equation} which one can write for $V(t)$ under the diffusion approximation. In this elementary account we shy away from stochastic calculus; the interested reader can consult a textbook such as  \cite{cm65,Gardiner:2004pv,Kampen:2007cr}. Eq.~\ref{eq:dVt} can be used to compute the moments of the input current, as shown next.

\subsubsection{The moments of the input current}
Recall that the synaptic input current of the LIF neuron is the term
\be \label{eq:Ilifapp}
I_i(t) = \sum_{j\neq i}^N J_{ij} \sum_k \delta(t-t_k^j),
\ee
and from Eq.~\ref{eq:V} we see that $I$ is related to the LIF neuron's membrane potential by
\be \label{eq:I-dotV}
I(t) = \dot V_t + {V_t \over \tau}.
\ee
This shows that the input current is related to the rate of change of $V$; correspondingly, the mean and variance of $I$ will be related to the rate of change of the mean and variance of $V$.\footnote{Notice how the variance of the input current of the binary neuron, Eq.~\ref{eq:sigma_binary}, also has units of variance over time; more on this later.} The rate of change of the mean and variance of $V$ are easy to compute using Eq.~\ref{eq:dVt} or Eqs.~\ref{eq:EdV}:
\be \label{eq:Vmom}
\lim_{dt \to 0} {\mathbb  E(dV_t) \over dt} = - { V_t \over \tau} + \tilde \mu, \quad \lim_{dt \to 0} {\text{Var}(dV_t) \over dt} = \tilde \sigma^2.
\ee
From this, using \ref{eq:I-dotV} and recalling that our results are conditioned on $V_t$ being fixed, we obtain
\be \label{eq:EI}
\mathbb E[I(t)] = \mathbb E \left [\dot V_t + {V_t \over \tau} \right ] = - { V_t \over \tau} + \tilde \mu + { V_t \over \tau} = \tilde \mu.
\ee
For the variance we get $\tilde \sigma^2$, since $V_t/\tau$ is fixed. In summary, recalling our definitions \ref{eq:momapp}, we have
\be \label{eq:Imom}
\mathbb E(I_t) = \tilde \mu = cN[J] f, \quad \text{Var}(I_t) = \tilde \sigma^2 = c N [J^2] f.
\ee
Adding up the contributions from distinct populations, we obtain relations that are identical to Eqs.~\ref{eq:c-mu-sigma} valid for binary neurons (the external current will pop up from $\mathbb E[\dot V_t]$ after redefinition of $V_t$, see footnote~\ref{ft:Vt}). This shows that Eq.~\ref{eq:varJx} (and hence the second of Eqs.~\ref{eq:c-mu-sigma}) is exact for integrate and fire neurons. The reason is the following: the discrete time step in the binary model is analogous to the elementary time step $dt$ in the LIF model ($x_j$ is really an increment); in the LIF model, the probability of spiking in bin $dt$ (analogous to $f$ in Eq.~\ref{eq:varJxexact}) is $f dt$, hence it is legitimate to neglect $f^2 dt^2$ as $dt \to 0$. Vice versa, in the binary neuron with discrete time step $dt$, Eq.~\ref{eq:varJx} would read $\text{Var}(J_{ij} x_j) \approx \mathbb E(J_{ij}^2) f_j dt$ to leading order in $dt$, or
\be
{\text{Var}(J_{ij} x_j) \over dt} \approx E(J_{ij}^2) f_j.
\ee
From this, we also recognize that Eq.~\ref{eq:varJx} is more properly understood as the rate of change of the variance over the duration of the discrete time step. This is also true of Eqs.~\ref{eq:Imom}, which technically are not the mean and variance of $I$ but their rate of change in continuous time (also known as {\em infinitesimal mean and variance} in the theory of diffusion processes). 

\vsp Although derived here in the context of LIF neurons, the result Eqs.~\ref{eq:Imom} holds in general, as the definition of the input current $I = \sum_j^N J_{ij} \sum_k \delta(t -t_k^j)$ does not depend on the neuron model (whilst the moments of $V$ depend on both $I$ and the specific neuron model). 

\newpage
\section*{}
\addcontentsline{toc}{section}{References}
\bibliographystyle{namedplus}

\small{
\bibliography{refs}

\end{document}